\documentclass[aps,preprint,showkeys,showpacs]{revtex4}%
\usepackage{amsfonts}
\usepackage{subfig}
\usepackage{amsmath}
\usepackage{amssymb}
\usepackage{graphicx}
\usepackage{epsfig}
\usepackage{color}

\begin{document}
\title{Dynamics of quantum Fisher information in a squeezed thermal bath}

\author{Asghar Ullah}
\affiliation{Department of Physics, Quaid-i-Azam University, Islamabad 45320, Pakistan.}
\begin{abstract}
	In this paper, the dynamics of quantum Fisher information of a qubit interacting with
	a squeezed thermal environment are studied. The optimal initial state of the qubit,
	temperature of the environment, and the interaction time, which maximize quantum
	Fisher information are obtained. Based on the ohmicity of the environment, we compare the dynamics
	of quantum Fisher information in ohmic, sub-ohmic, and super-ohmic regimes of the environment. Moreover, it is shown that the precise estimation of parameters is robust against squeezing.
\end{abstract}

\maketitle
\section{Introduction}
Quantum Fisher Information (QFI) is an important concept of Quantum estimation theory (QET) \cite{Helstrom, Holevo}, a direct extension of the classical estimation theory \cite{Fisher}. In QET, it sets up a lower bound for the estimation of parameters, which are not directly accessible to measurements and potentially affect the dynamics of different physical quantities of interest \cite{Javed, Paris, Benedetti1, Benedetti2}. Besides QET, it has numerous uses in other fields based on the applications of quantum mechanics. For example, in quantum information theory, it is used as a tool for measuring non-Markovianity of environments, quantum entanglement of many-body systems, quantum phase transition and distinguishability of quantum states on Hilbert spaces \cite{Wooters, Song, Hauke, Braunstein, Lu}.

A frequent strategy for studying the dynamics of QFI and using it as a tool for either estimating some unknown parameters or other characteristics of the environment is to monitor the evolution of an open quantum probe. Generally, a small controlled quantum system, usually a qubit, is allowed to interact with a bigger uncontrolled system (environment). During the interaction between the two, the information about the unknown parameter of the environment is encoded onto the space of the qubit. This encoded piece of information is then revealed through studying the dynamics of QFI which, in turn, enables for employing the estimation procedure. This strategy is very successful in studying the behavior of QFI under various setups. For instance, its dynamics provides useful information about the effects of single bosonic environment \cite{Lu, Zhong}, multiple bosonic environments \cite{Wang, Yan} and about the cutoff frequency of Ohmic type bosonic environments \cite{Benedetti3}.

In quantum thermodynamics, QFI has a pivotal role in estimating optimal temperature for different quantum mechanical processes \cite{25,26,27,28}. Many recent studies have focused on the measurement of temperature at very small scales in which nano-sized thermal environments are considered. The temperature of such environments become very sensitive to the disturbances caused by the presence of quantum probes \cite{31,32,33,34,35}. Quantum harmonic oscillators \cite{36}, atomic condensates \cite{37,38} and nanomechanical resonators \cite{39} are some interesting paradigms of nanoscale thermometry. In modern science and technology, precise estimation of optimal temperature has great significance on experimental realizations of many devices \cite{40,41,42}. 

In this paper, we study the dynamics of QFI as a function of temperature, time of interaction, squeezing amplitude and phase parameter for an exactly solvable model of a squeezed thermal environment interacting with a single qubit. In addition to the effect of temperature and time of interaction, we also investigate the effects of squeezing strength and relative phase of the environment on its behavior. Our results show that QFI can be maximized for certain values of both temperature of the environment and interaction time where the maximum varies with the choices of squeezing and phase parameters of the environment. Our findings, regarding the maximization of QFI with respect to temperature,  time of interaction, squeezing amplitude and phase parameter can be used to characterize the squeezed reservoir for practical purposes in the field of quantum metrology.

The paper is organized as follows: In section \ref{secII}, we introduce the Hamiltonian describing the dynamics of our system and briefly review the measure of QFI. In section \ref{secIII} we present our analytical results, followed by their numerical simulations and the relevant discussion. We summarize our work in section \ref{secIV}.
	
\section{Physical model}\label{secII}

The interaction of qubit with environment results in the creation of correlations between the states of the qubit and environment. This generally gives rise to dephasing and, as a result, decoherence, due to the interaction between the system and the various modes of the environment and, it is responsible for the destruction of quantum parallelism. This technique may be utilized effectively to make the qubit an effective probe to study various parameters of the environment, without compromising the energy of both systems. In this paper, the qubit-environment interaction is considered, i.e. a pure dephasing. The total Hamiltonian of the composite system may be expressed as

\begin{equation}
\hat{H}=\frac{1}{2}\omega_0\hat{\sigma}_z+ \sum_k\omega_k\hat{b}^\dagger_k \hat{b}_k+\hat{\sigma}_z.\sum_k(g_k\hat{b}^\dagger_k+g^*_k\hat{b}_k),\label{eq1}
\end{equation}
where the first part represents the qubit Hamiltonian, second part denotes the reservoir Hamiltonian and the last part indicates their interaction, $\omega_0$ ($\hbar=1$) is the transition frequency of the qubit, $\hat{\sigma}_z$ is the Pauli matrix that acts on the qubit. The frequency of $k$th modes of the reservoir is denoted by $\omega_k$ with coupling strength $g_k$ and $\hat{b_k}$ ($\hat{b^\dagger_k}$) is the annihilation (creation) bosonic operators. The commutation $[\hat{H}_S,\hat{H}_I]=0$ corresponds to the decoherence of qubit state.

The squeezed state of the reservoir is given as
\begin{equation}
\rho_R(0)=S_\xi\rho_{Th}S^\dagger_\xi,\label{eq2}
\end{equation}
where $\rho_{Th}=\exp[-\hat{H}_R/T]/Z$ is the thermal state (with $\hbar=1=k_B$), $Z=\text{Tr}(\exp[-\hat{H}_R/T])$ is the partition function, $\hat{H}_R$ is the reservoir Hamiltonian and $T$ is the temperature of reservoir. The squeezed operator $S_\xi$ in Eq.(\ref{eq2}) can be expressed as
\begin{equation}
S_{\xi}=\exp{\big[\frac{1}{2}(\xi^*\hat{b}^2_k-\frac{1}{2}\xi_k(\hat{b}^\dagger_k)^2\big)]},\label{eq3}
\end{equation}
where $\xi_k=r_k\exp[i\theta_k]$ containing the squeezing $r_k\ge0$, and phase parameter $\theta_k\in [0, 2\pi]$.
The time evolution of density matrix (DM) can be expressed as
\begin{equation}
\rho_S(t)=\text{Tr}_R\big[\hat{U}_I(t)\rho_{SR}(0)\hat{U}^{\dagger}_I(t)\big],\label{eq4}
\end{equation}
where $\rho_{SR}(0)=\rho_S(0)\otimes\rho_R(0)$ is the initial state of the composite system and $\rho_S(0)$ is the initial qubit state. In the interaction picture $\hat{U}(t)$ is given as 
\begin{equation}
\hat{U}_I(t)=\exp{\big[\frac{\hat{\sigma}_z}{2}\big(\alpha_k\hat{b}^\dagger_k-\alpha^*\hat{b}_k\big)\big]},\label{eq5}
\end{equation}
with $\alpha_k=2g_k(1-e^{i\omega_k t})/\omega_k$.

The QFI as a function of parameter $\eta$, for a spectrally decomposed DM is $\rho(\eta)=\sum_i\lambda_i(\eta)|\phi_i(\eta)\rangle\langle\phi_i(\eta)|$, can be written as \cite{Benedetti3}
\begin{equation}
I(\eta)=\sum_{i}\frac{({\partial_{\eta}\lambda_i(\eta)})^2}{\lambda_i(\eta)}+2\sum_{i\neq j}\frac{(\lambda_i(\eta)-\lambda_j(\eta))^2}{\lambda_i(\eta)+\lambda_j(\eta)}|\langle\phi_i(\eta)|\partial_{\eta}\phi_j(\eta)\rangle|^2,\label{eq6}
\end{equation}
where the first part is classical Fisher information (CFI)\cite{54} and, the second one is quantum mechanical in nature and represents QFI. The reduction of QFI to CFI corresponds to the optimal measurement. Since QFI is used as an estimation tool therefore, it is worthwhile to find the parameters which maximize QFI\cite{54,Javed}.

\section{Results and discussions}\label{secIII}

We present our results by employing the mathematical machinery developed in the previous section. We begin from a generic superposition of qubit state $|\psi(0)\rangle=\cos(\alpha/2)|0\rangle+\sin(\alpha/2)|1\rangle$. Using Eqs.(\ref{eq1}) and (\ref{eq3}) in Eq.(\ref{eq2}), combining the result with $\rho_S(0) =|\psi(0)\rangle\langle\psi(0)|$ for $\rho_{SR}(0)$ and putting it along with Eq.(\ref{eq5}) in Eq.(\ref{eq4}) leads to the reduced DM of qubit, which can be expressed as
\begin{equation}
\rho_s(t)=\begin{pmatrix}
\cos^2(\alpha/2) & \sin(\alpha/2)\cos(\alpha/2) \exp[-\Gamma(T,t)]\\
 \sin(\alpha/2)\cos(\alpha/2) \exp[-\Gamma(T,t)] &  \sin^2(\alpha/2)
\end{pmatrix}\label{eq7}
\end{equation}
where $\Gamma(T,t)$ carries the information of environment, stored in the the qubit state with its exponent, such as
\begin{align}
\exp[-\Gamma(T,t)]=\sum_k\langle \exp[\eta_k\hat{b}^\dagger-\eta^*\hat{b}_k]\rangle.\label{eq8}
\end{align}
This, in fact, is characterized by Wigner representation of $\rho_R(0)$, representing all modes of squeezed state \cite{Zhong,Benedetti3}. In terms of the environmental characteristic parameters, the function $\Gamma(T,t)$ can be written as
\begin{align}
\Gamma(T,t)=\sum_k\frac{1}{2}|\eta_k(t)|^2\coth\left(\frac{\omega_k}{2T}\right),\label{eq9}
\end{align}
with
\begin{align}
\eta_k(t)=\alpha_k\cosh r_k+\alpha^*_ke^{i\theta_k}\sinh r_k.\label{eq10}
\end{align}
For an environment of continuous mode distribution the summation over $|g_k|^2$ can be replaced with integral over the spectral density function $J(\omega)$ of the environmental modes, that is, $\sum_k|g_k|^2\rightarrow\int J(\omega) d\omega$. Using the value of $\alpha_k$ from Eq.(\ref{eq5}), it can be expressed as
\begin{align}
\Gamma(T,t)=\int J(\omega)\Big(\frac{1-\cos\omega t}{\omega^2}\Big)[\cosh2r-\cos(\theta-\omega t)\sinh2r]\coth\left(\frac{\omega}{2T}\right)d\omega.\label{eq11}
\end{align}
 For an environment belongs to ohmic family, the spectral density function is given by
\begin{align}
J(\omega,\omega_c)=\frac{\omega^s}{\omega^{s-1}_c}e^{-\omega/\omega_c},\label{eq12}
\end{align}
where a system frequency response boundary is given by cut-off frequency$(\omega_c)$, and $s$ is the ohmicity dimensionless parameter, which grades the reservoir in three classes, ohmic $(s=1)$, sub-ohmic $(s<1)$ and super-ohmic $(s>1)$.
The eigenvalues and eigenvectors of the final DM in Eq.(\ref{eq7}) is expressed as
\begin{align} 
\lambda^{\pm}(T,t)=\frac{1}{2}[1\pm e^{-\Gamma(T,t)}\chi(\Gamma,\alpha)],\label{eq13}
\end{align}
and
\begin{align} 
\phi^{\pm}(T,t)=\left[e^{\Gamma(T,t)}\,\cot\theta\pm \csc\theta\,\chi(\Gamma,\alpha)\right]|0\rangle+|1\rangle,\label{eq14}
\end{align}
with function $\chi(\Gamma,\alpha)=\sqrt{e^{2\Gamma(T,t)}\,\cos^2\alpha+\sin^2\alpha}$.\\ Substitution of Eqs. (\ref{eq13}) and (\ref{eq14}) into Eq.(\ref{eq6}) leads to the following form for QFI
\begin{align} 
I(T,t)=\frac{\sin^2\alpha\,[\partial_T\Gamma(T,t)]^2}{\exp[2\Gamma(T,t)]-1},\label{eq15}
\end{align}
where, $\partial_T=\frac{\partial}{\partial T}$. In Eq.(\ref{eq15}), the derivative is taken w.r.t the parameter that is to be estimated (in our case, these are $r$, $\theta$ and $T$).\\
It can be seen that the initial state parameter $\alpha$ maximizes Eq.(\ref{eq15}) for $\alpha=\pi/2$ , which corresponds to the equatorial state $|\psi(0)\rangle=|+\rangle=(|0\rangle+|1\rangle)/\sqrt{2}$ of the qubit. In other words, the maximal superposition of the computational bases is the optimal initial qubit state. With this choice of the initial qubit state, the function $\chi(\Gamma,\alpha)=1$ and the eigenfunctions in Eq.(\ref{eq14}) becomes independent of temperature $T$. This, in turn, reduces the second part of Eq.(\ref{eq6}) to zero, which means that the projectors corresponding to the computational bases of the qubit constitute the set of optimal measurement. Finally, we resort to numerical simulation of Eq.(\ref{eq15}) for the optimal initial qubit state due to non-vocational behavior of  $\Gamma(T,t)$.

\begin{figure}[!]
	\centering
	\subfloat[]{
		\includegraphics[scale=0.3]{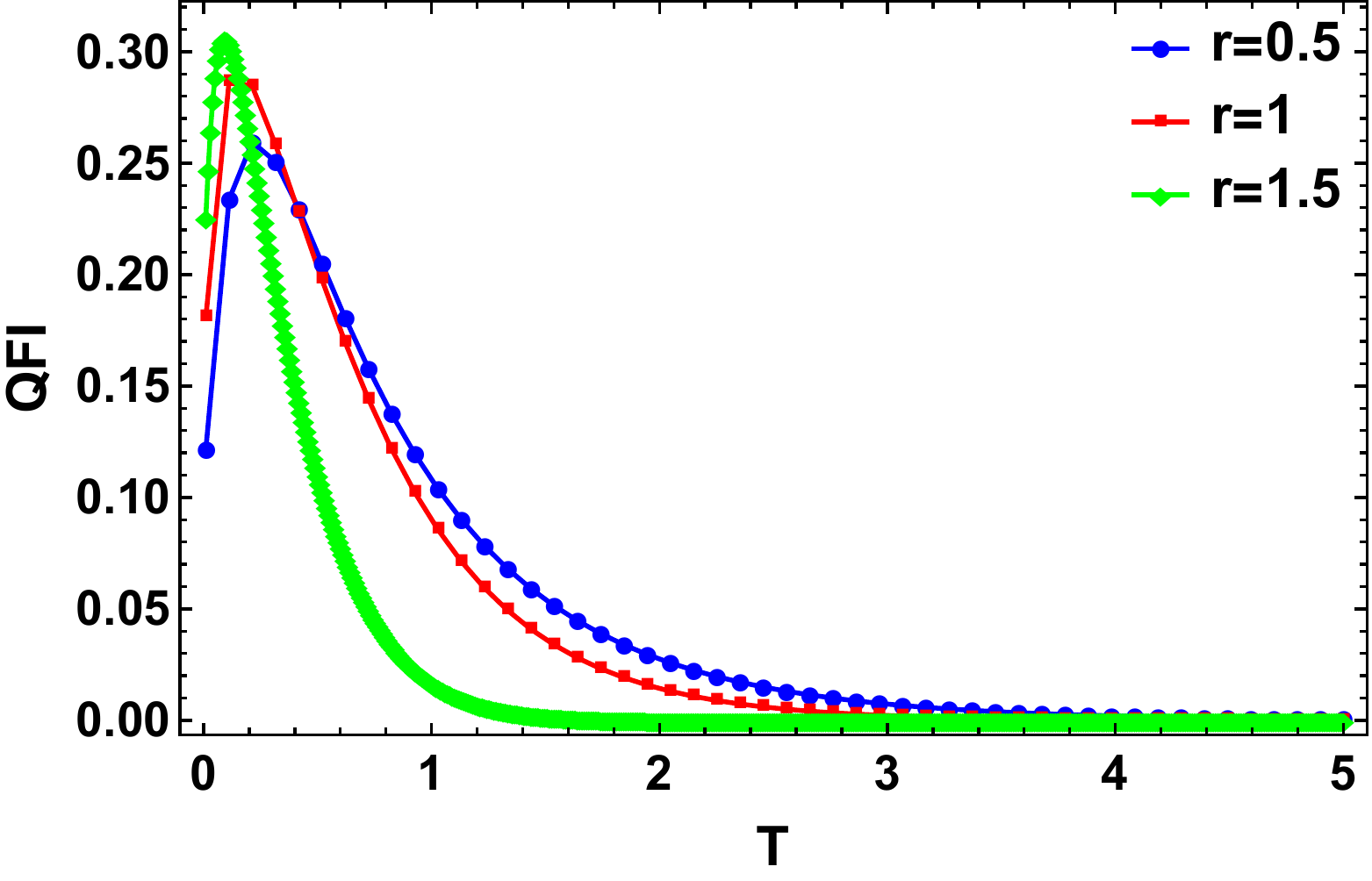}\label{Fig1a}}\hspace{0.4cm}
	\subfloat[]{
		\includegraphics[scale=0.3]{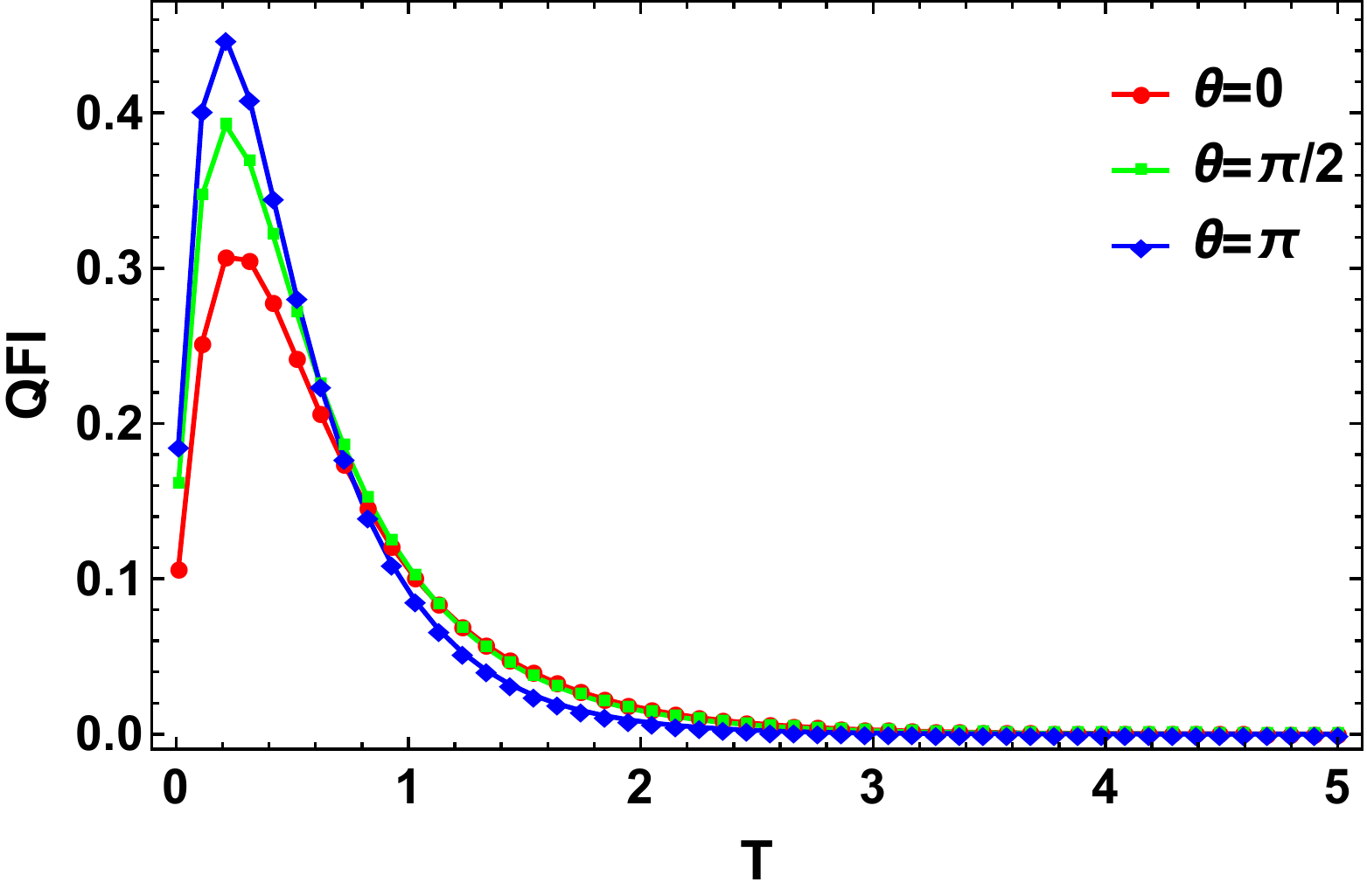}\label{Fig1b}}\\
	\subfloat[]{
		\includegraphics[scale=0.3]{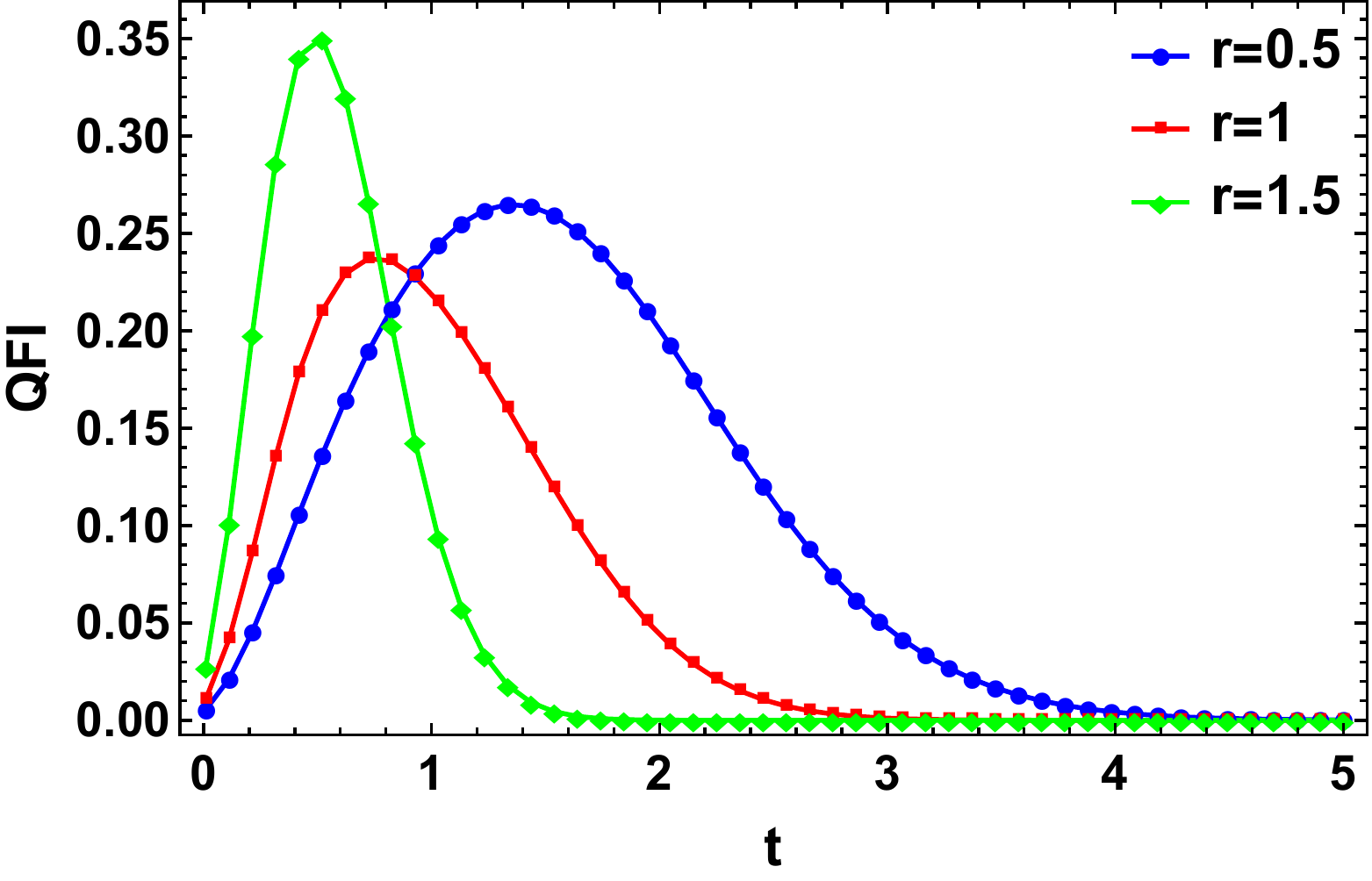}\label{Fig1c}}\hspace{0.4cm}
	\subfloat[]{
		\includegraphics[scale=0.3]{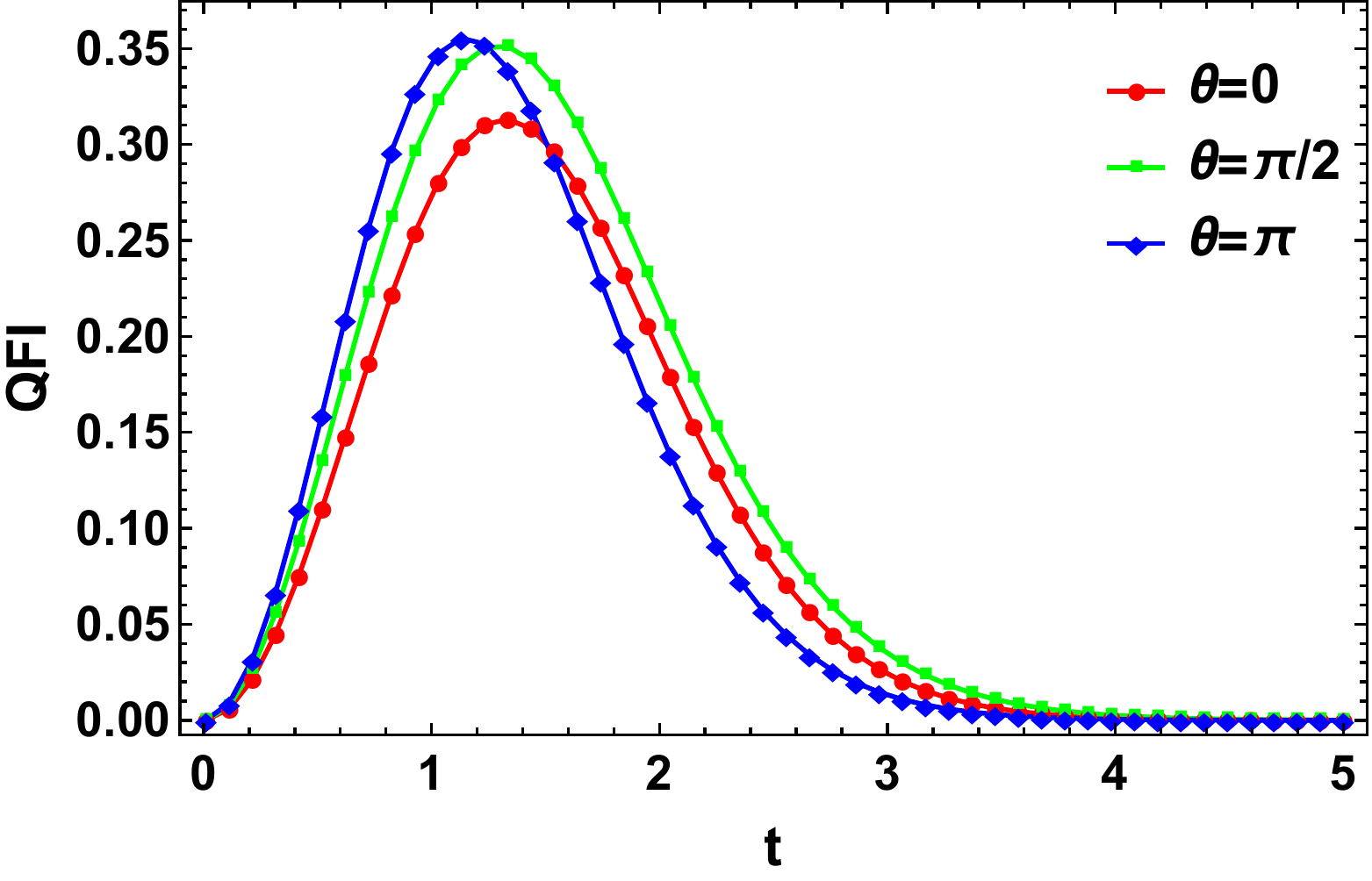}\label{Fig1d}}
	\caption{(Colour online) QFI verses temperature T (a,b) and QFI versus time t (c,d) for sub ohmic $(s=0.5)$ case. Here we used (a) $t=1$, $\theta=1$ (b) $t=1$, $r=0.1$ (c) $T=0.5$, $\theta=1$ and (d) $T=0.5$, $r=0.1$.}\label{fig1}
\end{figure}

 The dynamics of QFI as a function of temperature $T$ of the environment and the interaction time $t$ for three different values of the squeezing parameter $r$ and the phase parameter $\theta$ for the sub-ohmic regime with $s=0.5$ is plotted in Figure \ref{fig1}. The qualitative behavior of QFI is not considerably affected by the choices of the parameters $r$ and $\theta$ when plotted as a function of temperature. In each case, it first rapidly grows to a maximum and then monotonically falls to a vanishing value with increasing temperature. The peak shifts minutely to lower temperature with increasing $r$ (Figure (\ref{Fig1a})).  Also, in both cases, the peak value and the rate of decrease, beyond the maximum, varies as per the choice of $r$ and $\theta$. The peak values are highest for $r=1.5$ and $\theta=\pi$ in the two Figures, respectively. On the other hand, a gradual rate of increase and decrease at both sides of the peak is observed and the peaks become broadened when plotted as a function of the time of interaction. In fact, maximum information about the environment is stored in the space of the qubit at the peaks and thus they enhance the optimality of the projective measurements over the qubit state. This kind of behavior of QFI is very beneficial in various estimation procedures \cite{Zhong, Javed}.

\begin{figure}[!]
	\centering
	\subfloat[]{
		\includegraphics[scale=0.3]{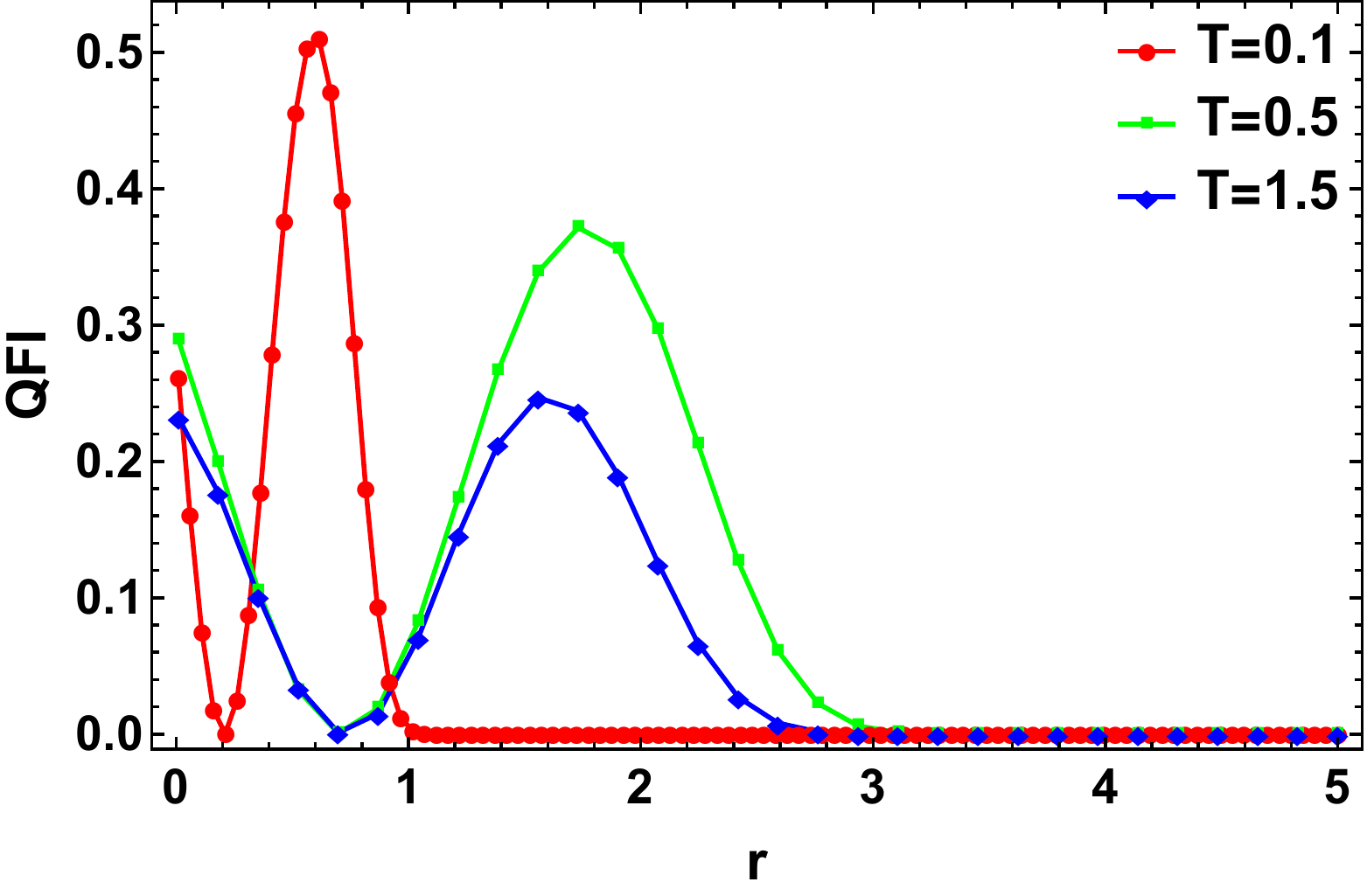}\label{Fig2a}}\hspace{0.4cm}
	\subfloat[]{
		\includegraphics[scale=0.3]{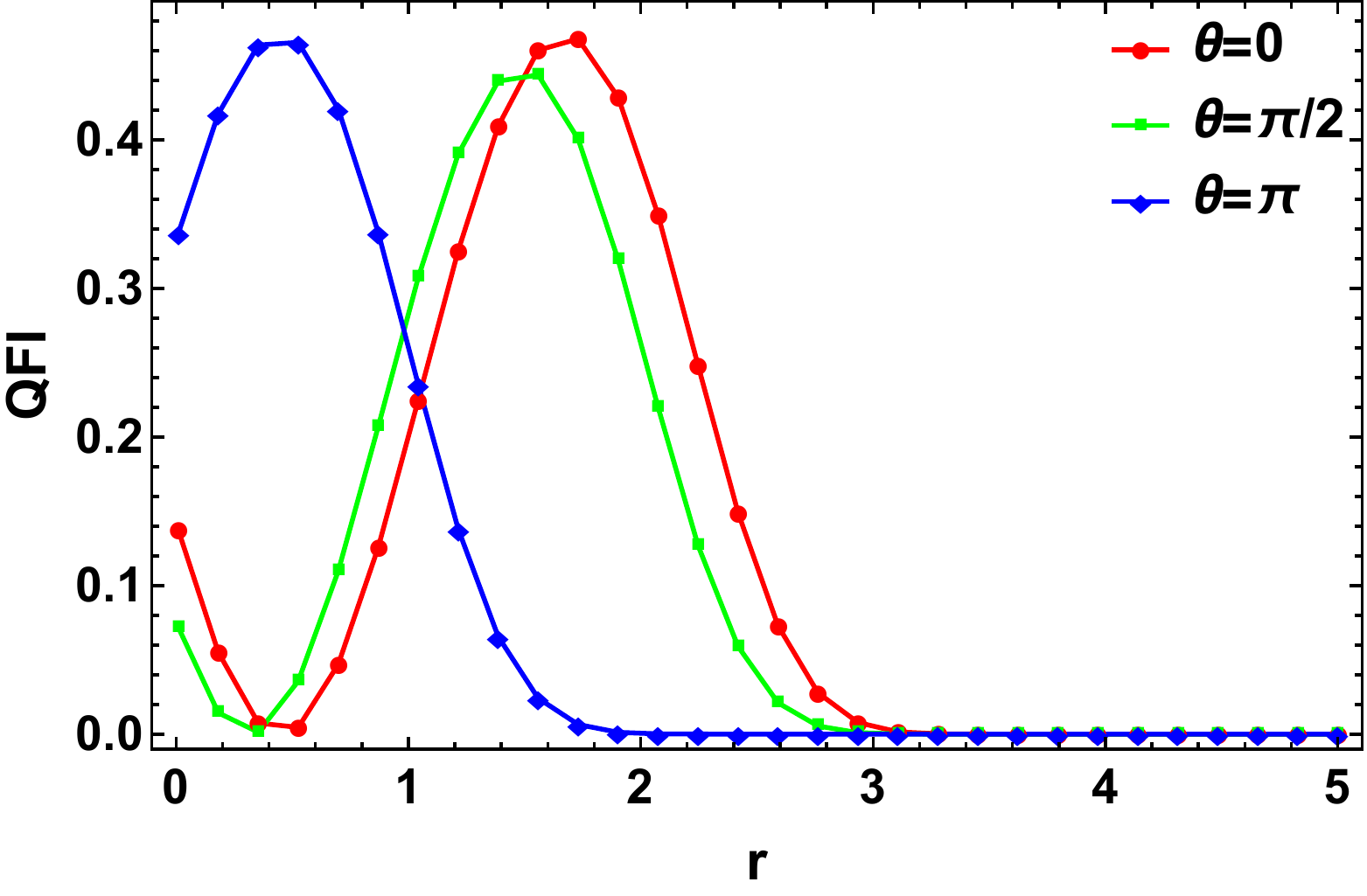}\label{Fig2b}}\\
	\subfloat[]{
		\includegraphics[scale=0.3]{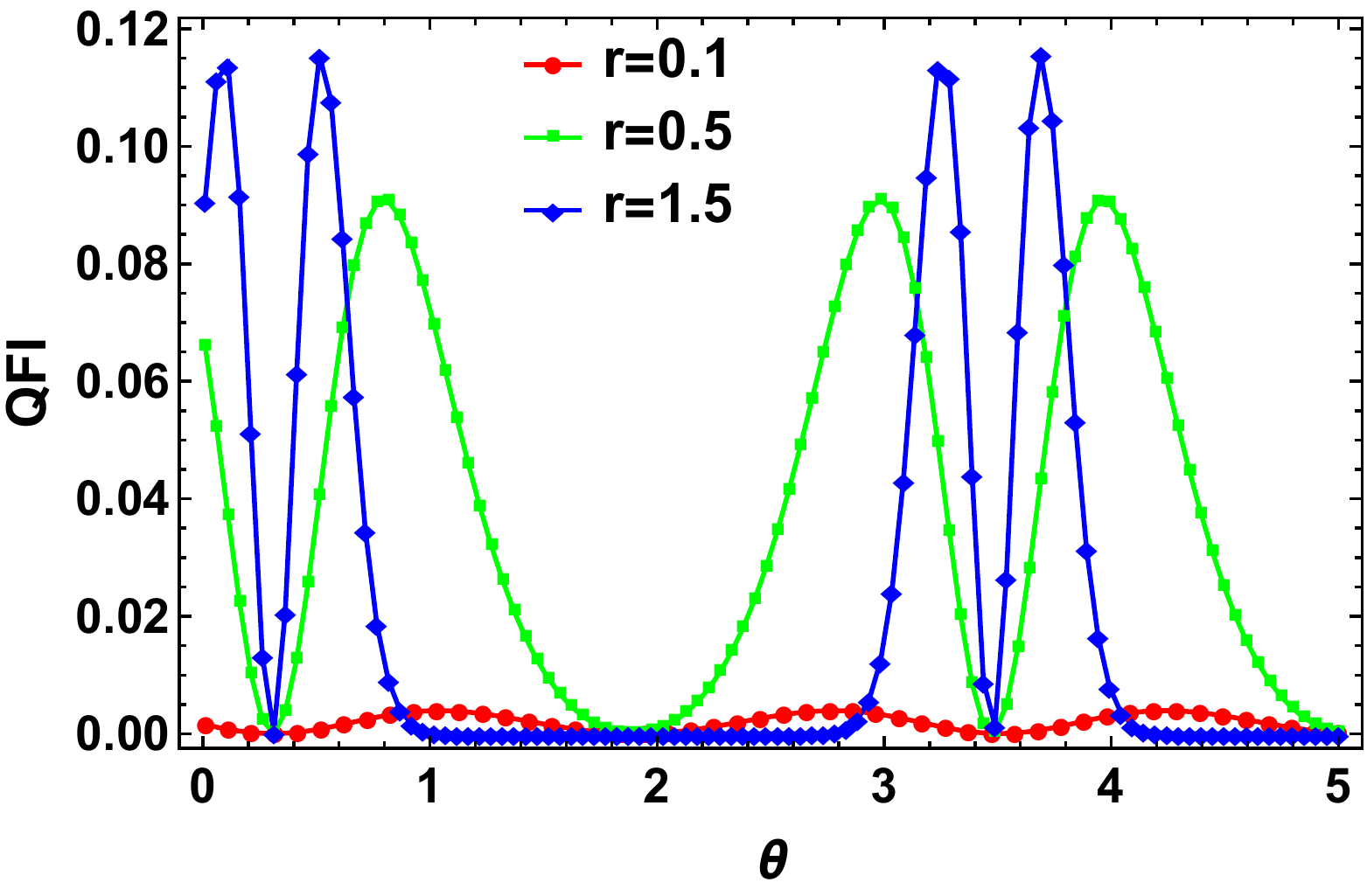}\label{Fig2c}}\hspace{0.4cm}
	\subfloat[]{
		\includegraphics[scale=0.3]{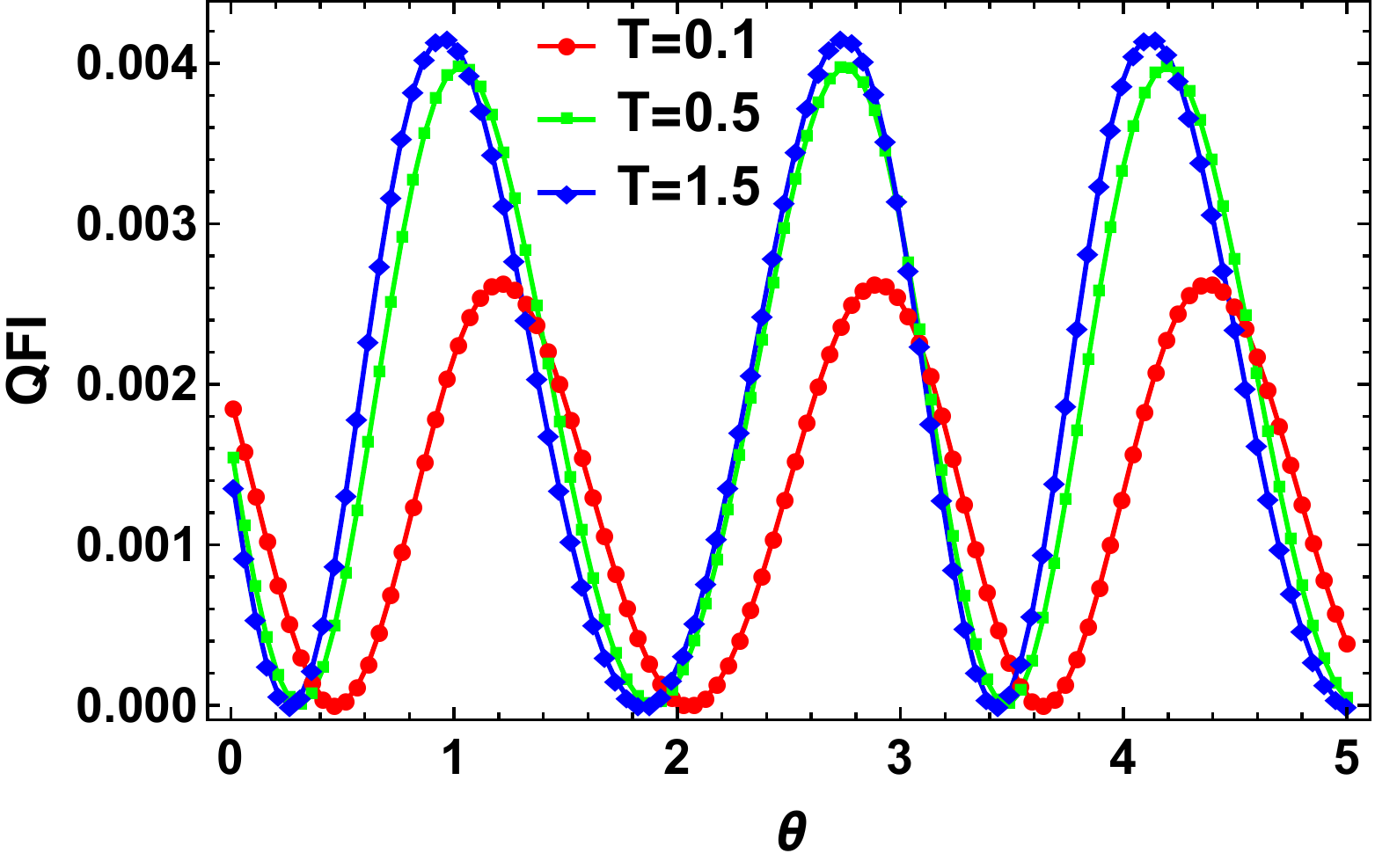}\label{Fig2d}}
	\caption{(Colour online) QFI verses squeezing strength $r$ (a,b) and QFI versus phase parameter $\theta$ (c,d) for sub ohmic $(s=0.5)$ case. Here we used (a) $t=1$, $\theta=1$ (b) $t=1$, $T=0.5$ (c) $T=0.5$, $t=1$ and (d) $t=1$, $r=0.1$.}\label{Fig2}
\end{figure}

Now we examine the dynamical behavior of QFI against squeezing strength $r$ (upper panel) and phase $\theta$ (lower panel) for different values of $T$ , $\theta$ and, $r$, $T$ respectively for sub-ohmic regime ($s=0.5$). As shown in Figure \ref{Fig2}, at low temperature i.e. $T=0.1$, the QFI is maximum but at high temperature, the peak values are reduced and becomes broadened which reduces QFI. The peak values are not affected notably by different choices of $\theta$, it only shifts the peak towards the left at $\theta=\pi$ (Figure \ref{Fig2b}) which reduces the time needed to extract information about $r$ from the environment. Periodic behavior of QFI is observed when plotted with respect to $\theta$. Squeezing parameter $r$ greatly influences the value of QFI and, it becomes maximum at $r=1.5$. This result indicates that squeezed fields may yield more information about the phase parameter $\theta$. However, an increase in temperature has no noteworthy effect on QFI and it minutely increases the peak value and the peaks roughly become the same at higher values of T and, once again a periodic behavior of QFI is observed (Figure \ref{Fig2d}).

\begin{figure}[!]
	\centering
	\subfloat[]{
		\includegraphics[scale=0.3]{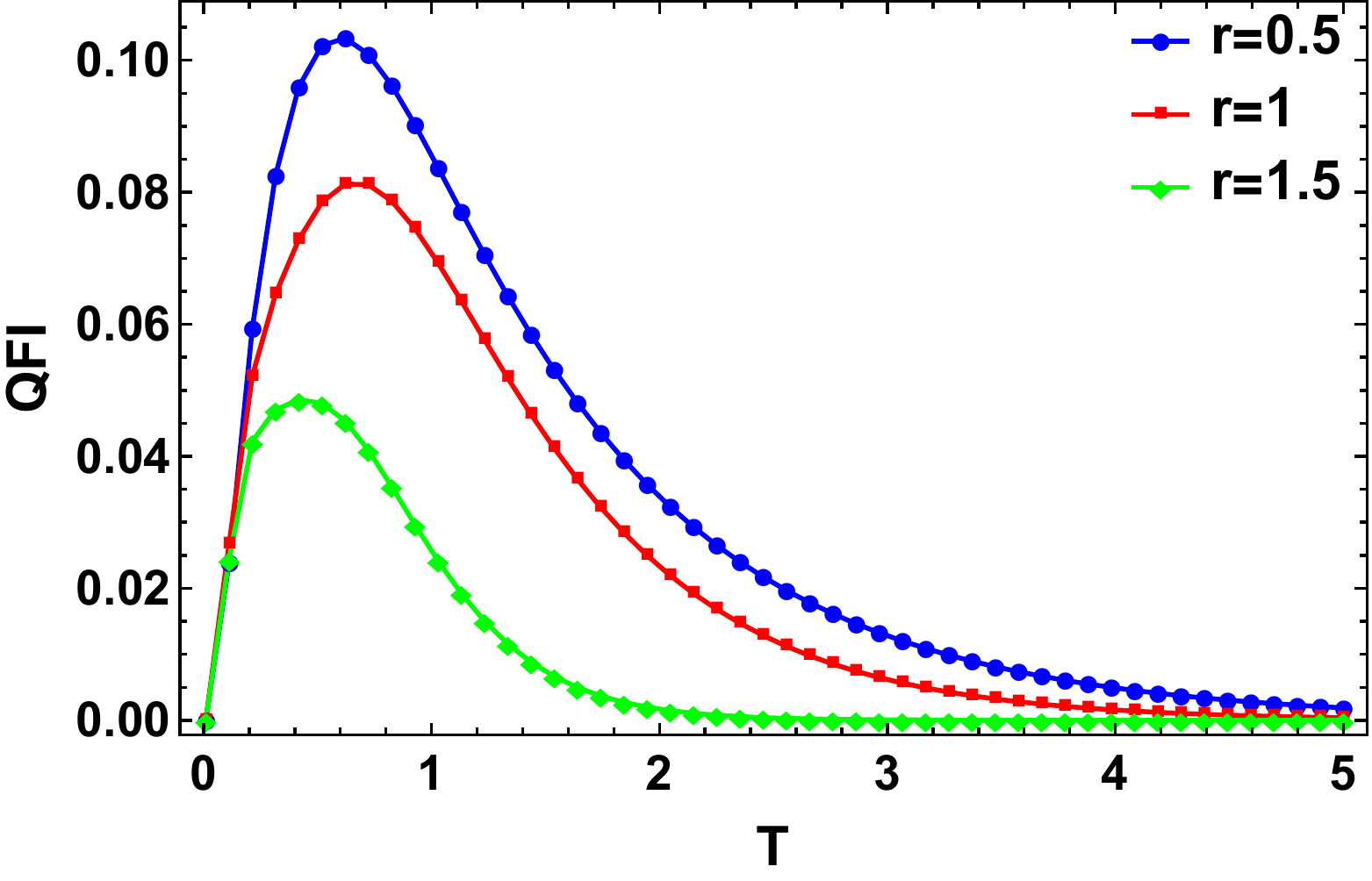}\label{Fig3a}}\hspace{0.4cm}
	\subfloat[]{
		\includegraphics[scale=0.3]{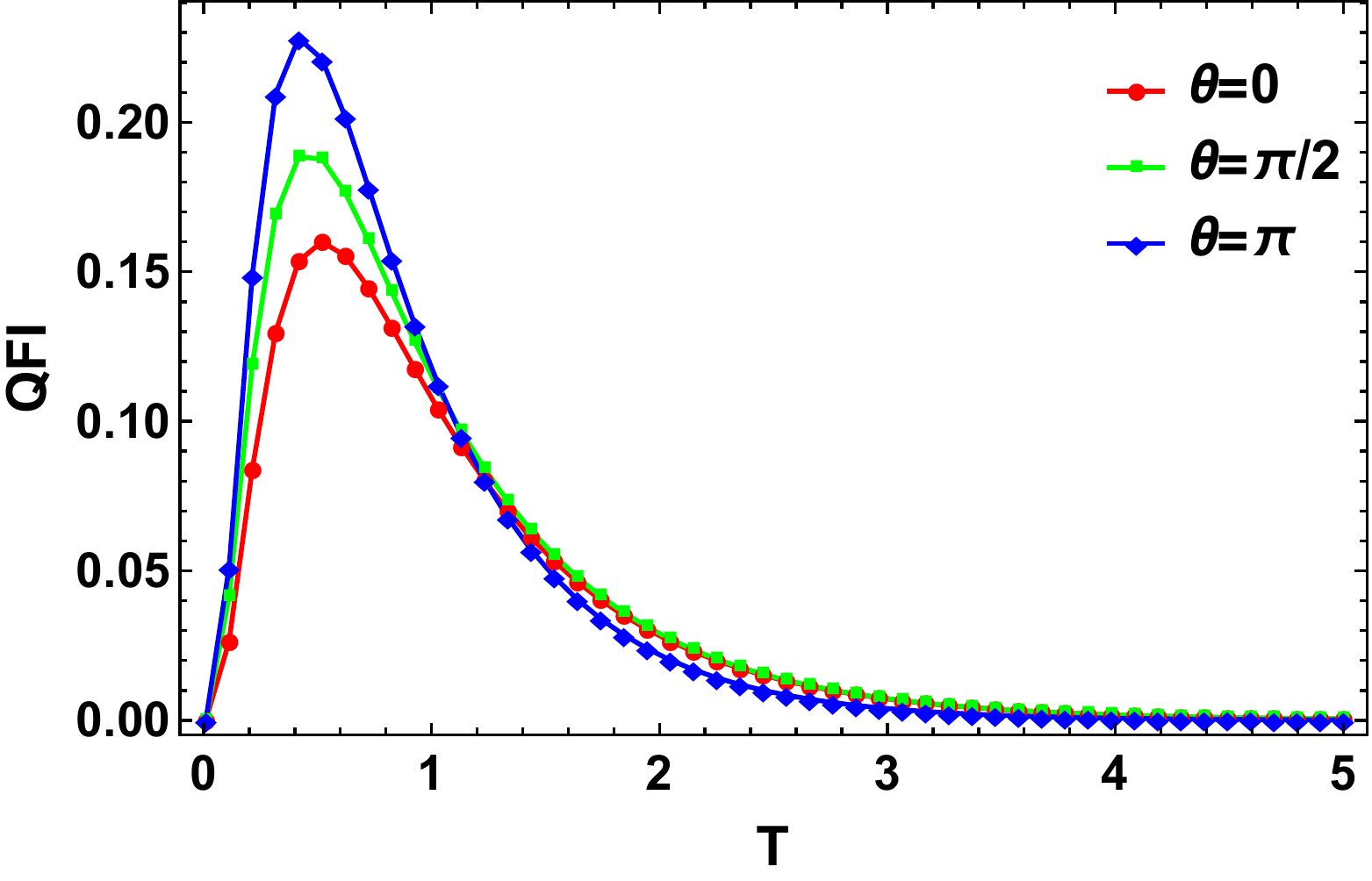}\label{Fig3b}}\\
	\subfloat[]{
		\includegraphics[scale=0.3]{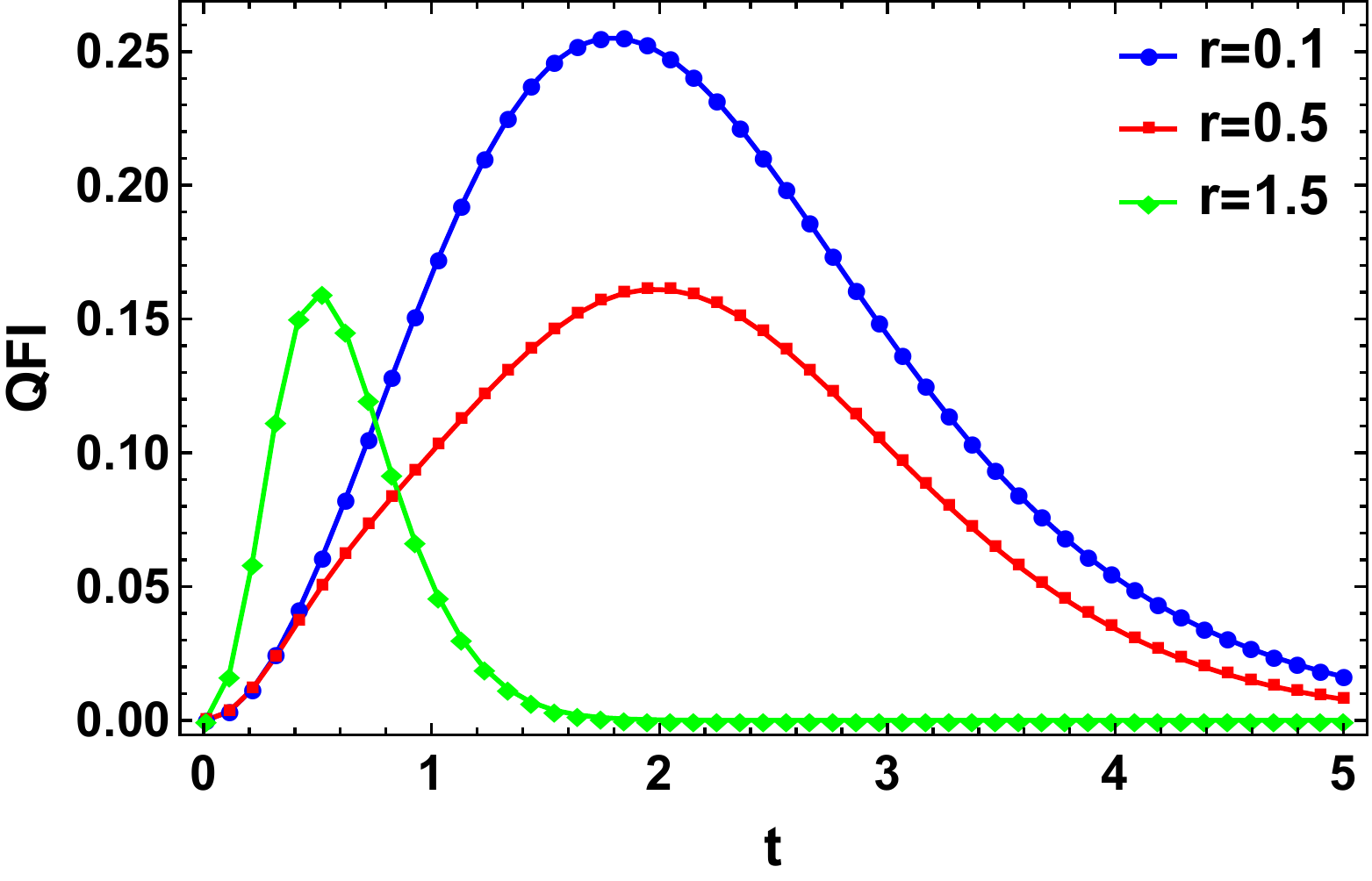}\label{Fig3c}}\hspace{0.4cm}
	\subfloat[]{
		\includegraphics[scale=0.3]{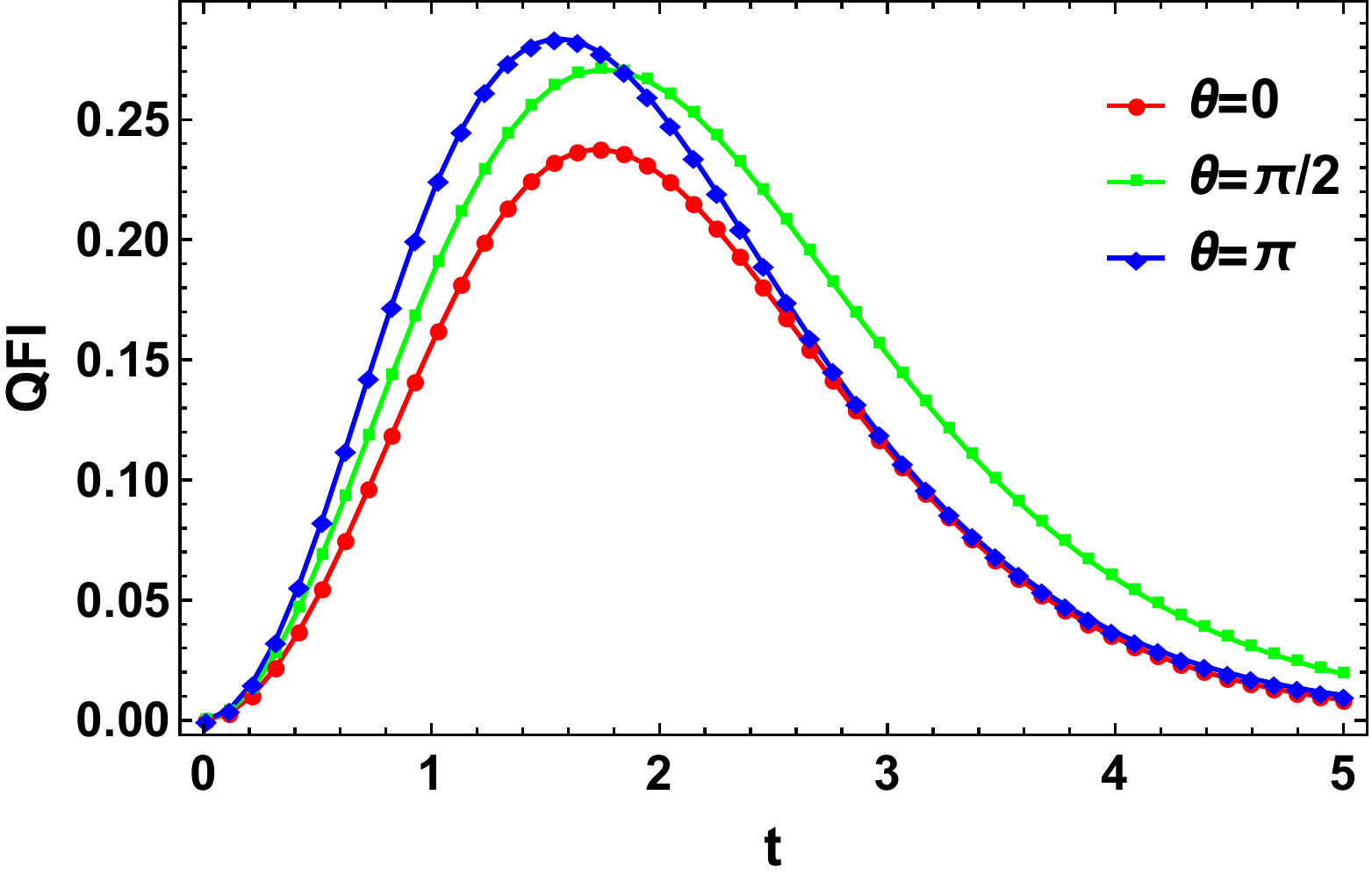}\label{Fig3d}}
	\caption{(Colour online) QFI verses temperature T (a,b) and QFI versus time t (c,d) for ohmic $(s=1)$ case.  Here we used (a) $t=1$, $\theta=1$ (b) $t=1$, $r=0.1$ (c) $T=0.5$, $\theta=1$ and (d) $T=0.5$, $r=0.1$.}\label{fig3}
\end{figure}

The behavior of QFI in the ohmic case $(s=1)$ of the environment for the same values of parameters as in Figure \ref{fig1} is shown in Figure \ref{fig3}. One can recognize that in each case the peak values are highly reduced as compared to the case of the sub-ohmic regime, however, the peaks are relatively broadened and shifted to the right in each case. On one hand, the reduced values of the peaks confine the utility of QFI both in quantum estimation procedures and quantum information theory. On the other hand, the broadened peaks help in increasing the precision of estimability in different quantum estimation processes. The shifting of peaks to the regime of higher temperature increases the easiness of its experimental realizations. Another salient feature, in comparison to Figure \ref{fig1}, is the order of the peaks which is reversed with respect to the values of the squeezing parameter $r$.

\begin{figure}[!]
	\centering
	\subfloat[]{
		\includegraphics[scale=0.3]{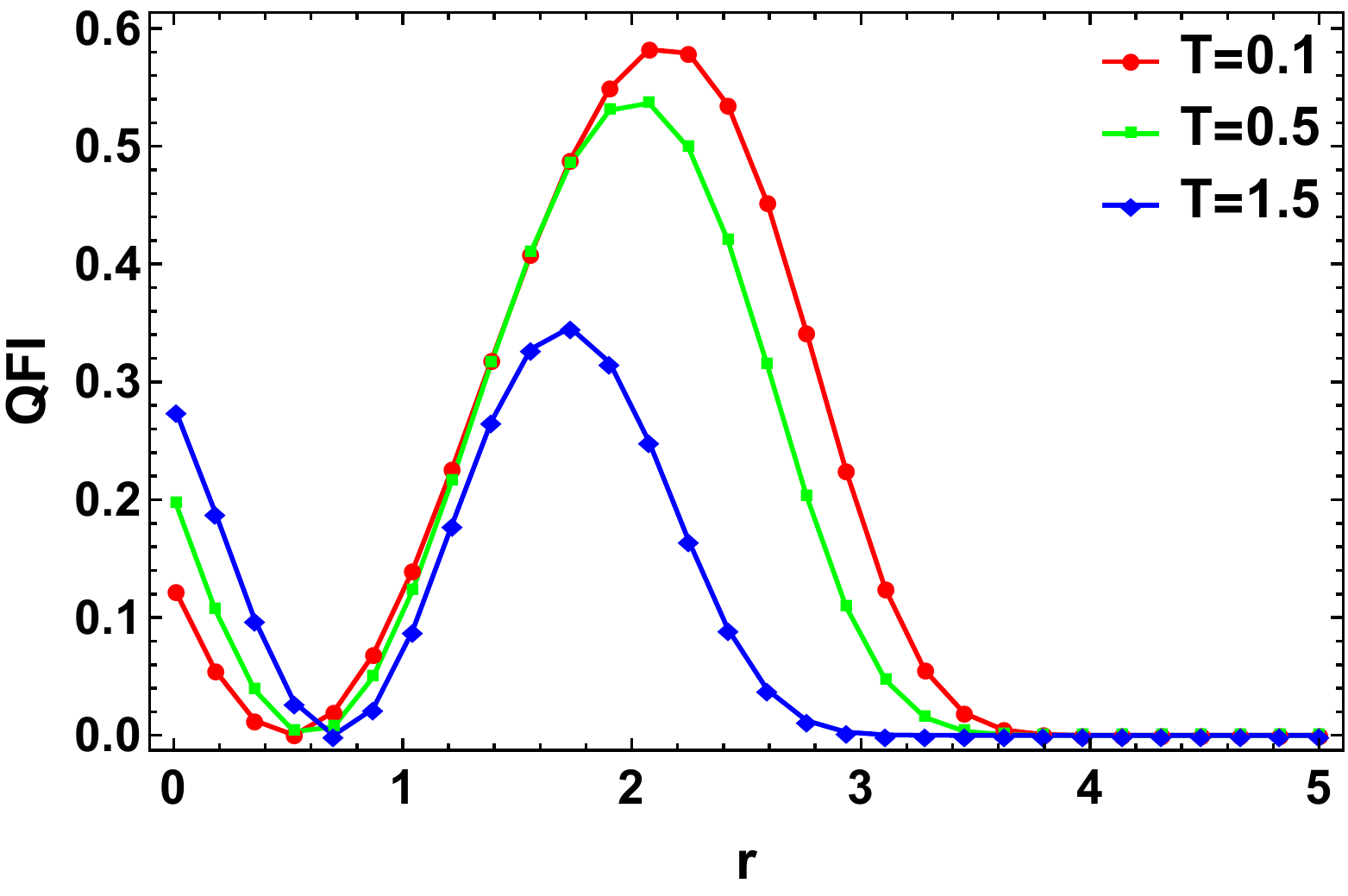}\label{Fig4a}}\hspace{0.4cm}
	\subfloat[]{
		\includegraphics[scale=0.3]{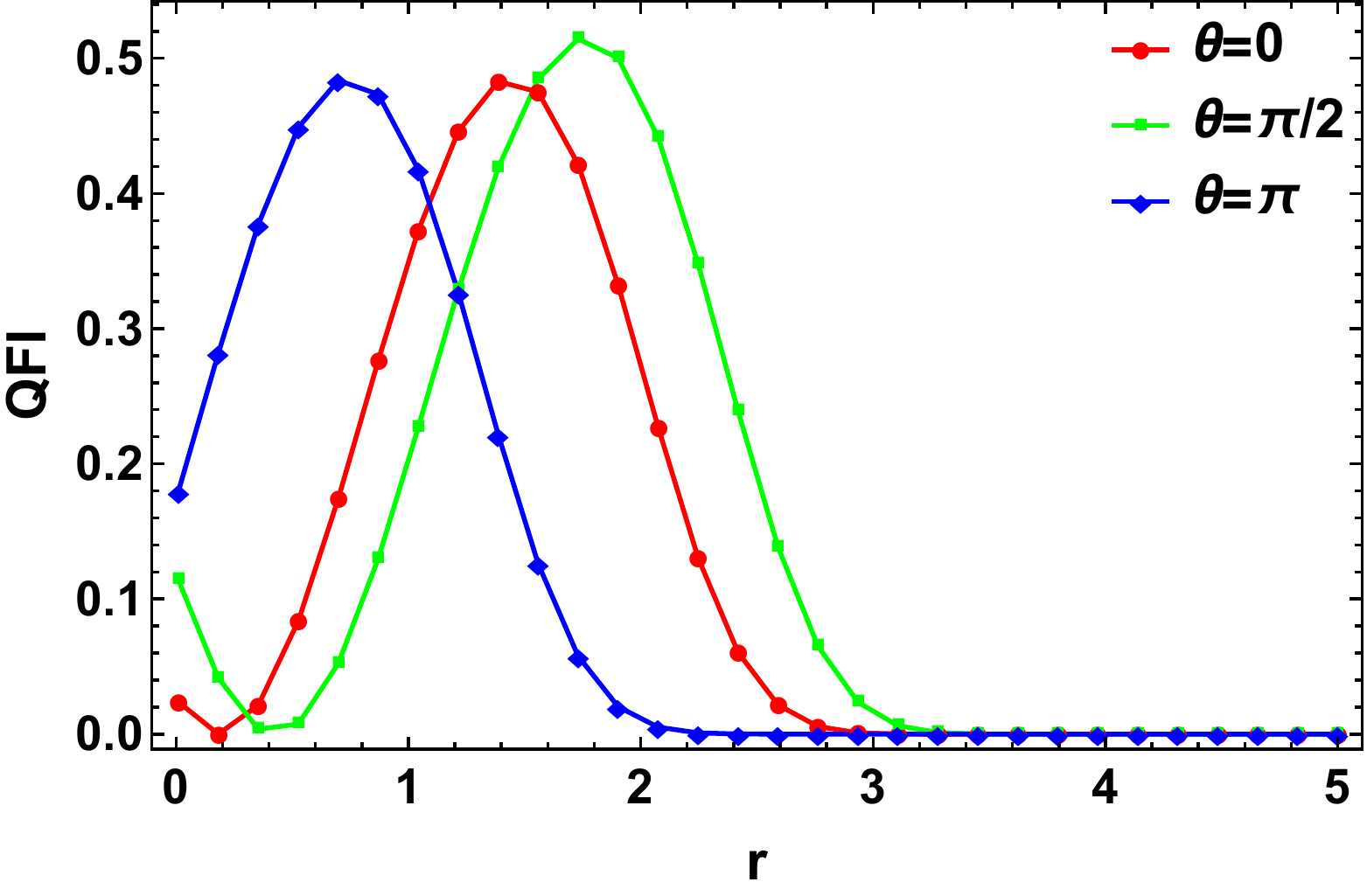}\label{Fig4b}}\\
	\subfloat[]{
		\includegraphics[scale=0.3]{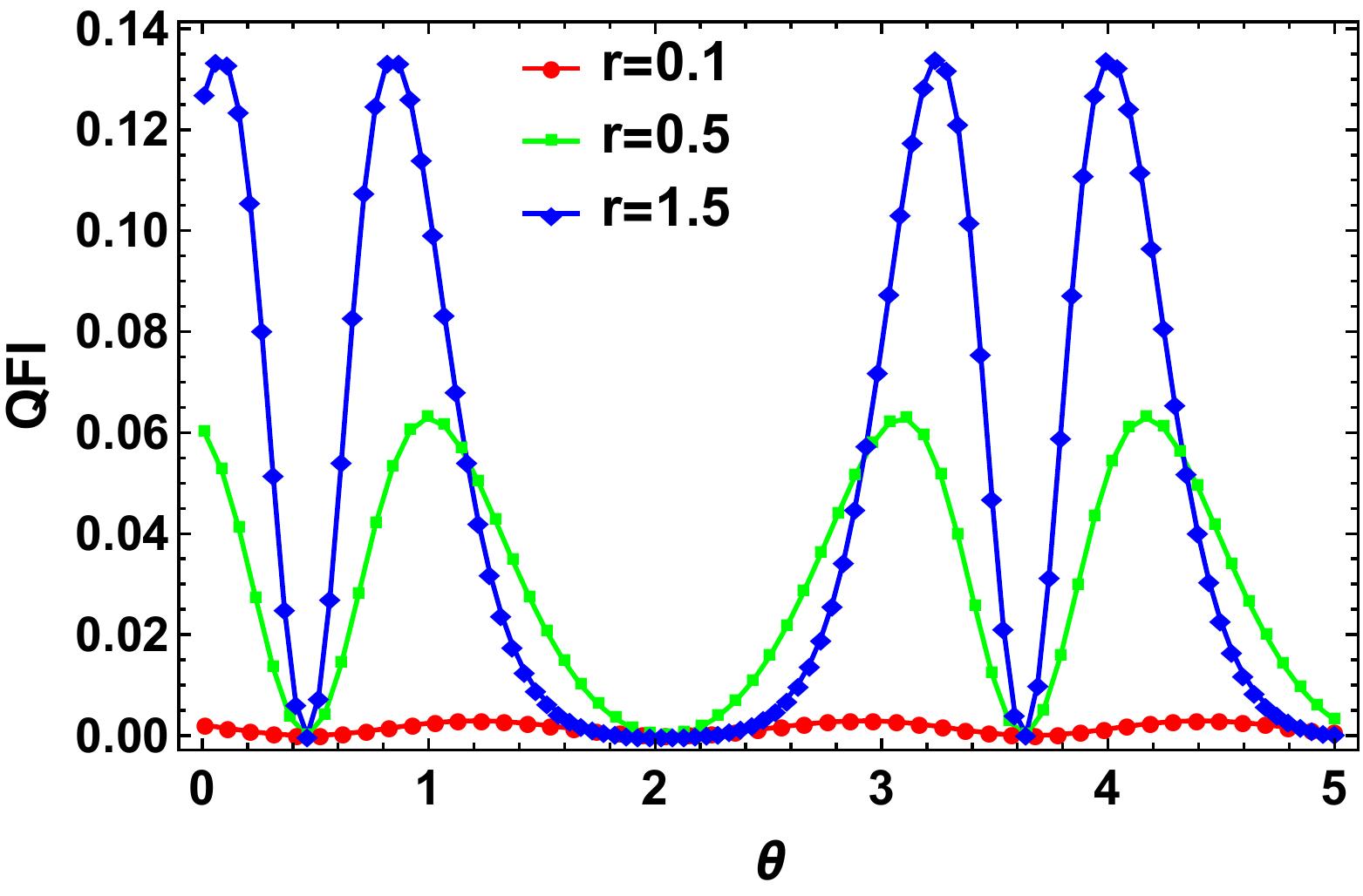}\label{Fig4c}}\hspace{0.4cm}
	\subfloat[]{
		\includegraphics[scale=0.3]{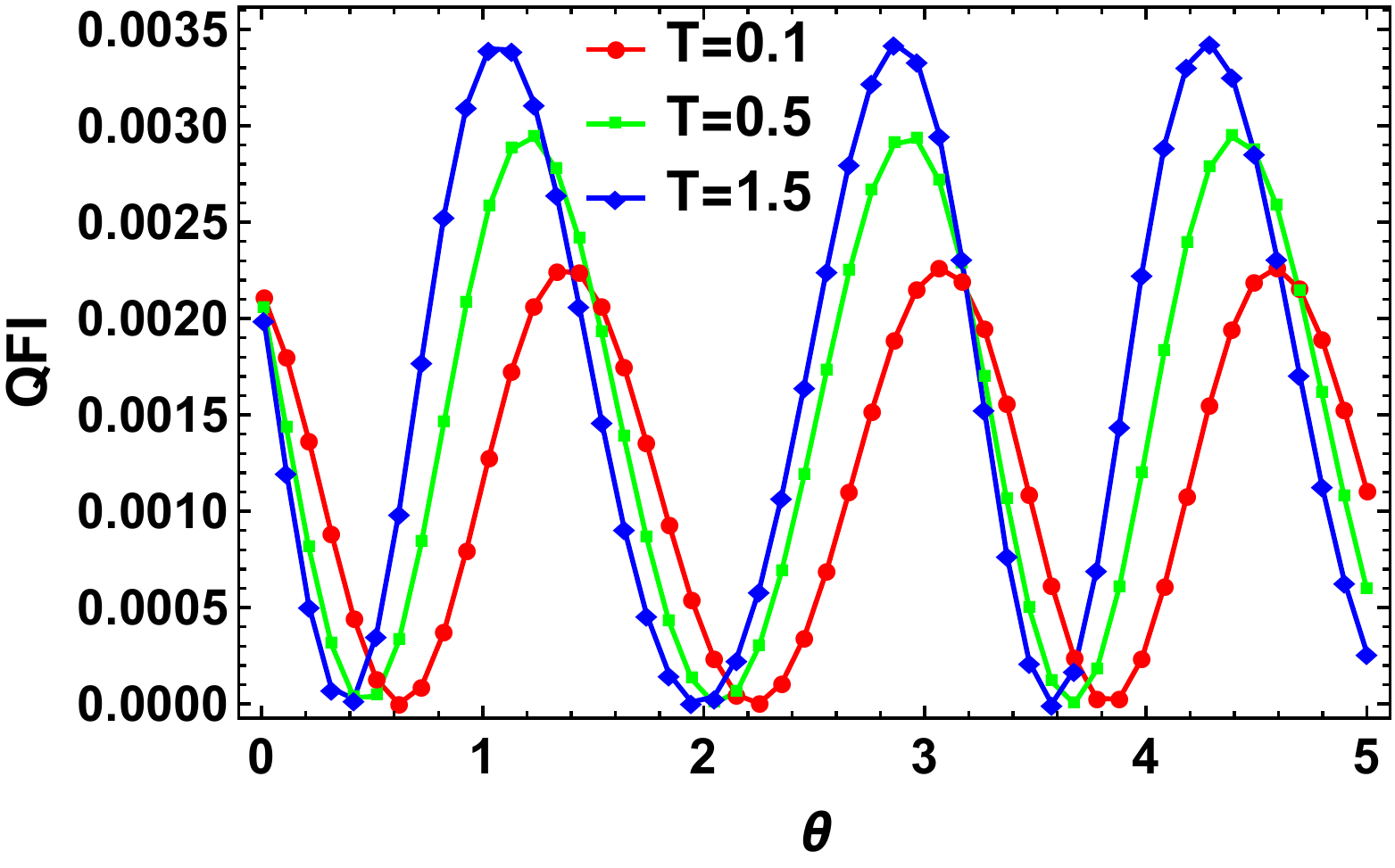}\label{Fig4d}}
	\caption{(Colour online) QFI verses squeezing strength $r$ (a,b) and QFI versus phase parameter $\theta$ (c,d) for ohmic $(s=1)$ case. Here we used (a) $t=1$, $\theta=1$ (b) $t=1$, $T=0.5$ (c) $T=0.5$, $t=1$ and (d) $t=1$, $r=0.1$.}\label{Fig4}
\end{figure}

Figure \ref{Fig4}, describes the dynamics of QFI as a function of $r$ and $\theta$ for ohmic regime $(s=1)$ for the same set of values of rest parameters as shown in Figure \ref{Fig2}. The peak values of QFI decrease as T is increased from 0.1 to 1.5 shown in Figure (\ref{Fig4a}). This illustrates that a high-temperature regime is not suitable to extract information about $r$ from the environment. As noticed in the sub-ohmic case, the phase parameter only shifts the peak towards left and, the maximum value of QFI is not affected remarkably with different choices of $\theta$ in Figure (\ref{Fig4b}). In Figure (\ref{Fig4c}), the QFI vanishes at $r=0.1$ however, the maximum value of QFI can be observed at $r=1.5$ with its periodic behavior. In the same aspect, QFI increase as $T$ is enhanced. From these observations, it is summarized that a low-temperature regime and squeezed environment offer suitability for the estimation of $r$ and $\theta$ respectively.

\begin{figure}[!]
	\centering
	\subfloat[]{
		\includegraphics[scale=0.25]{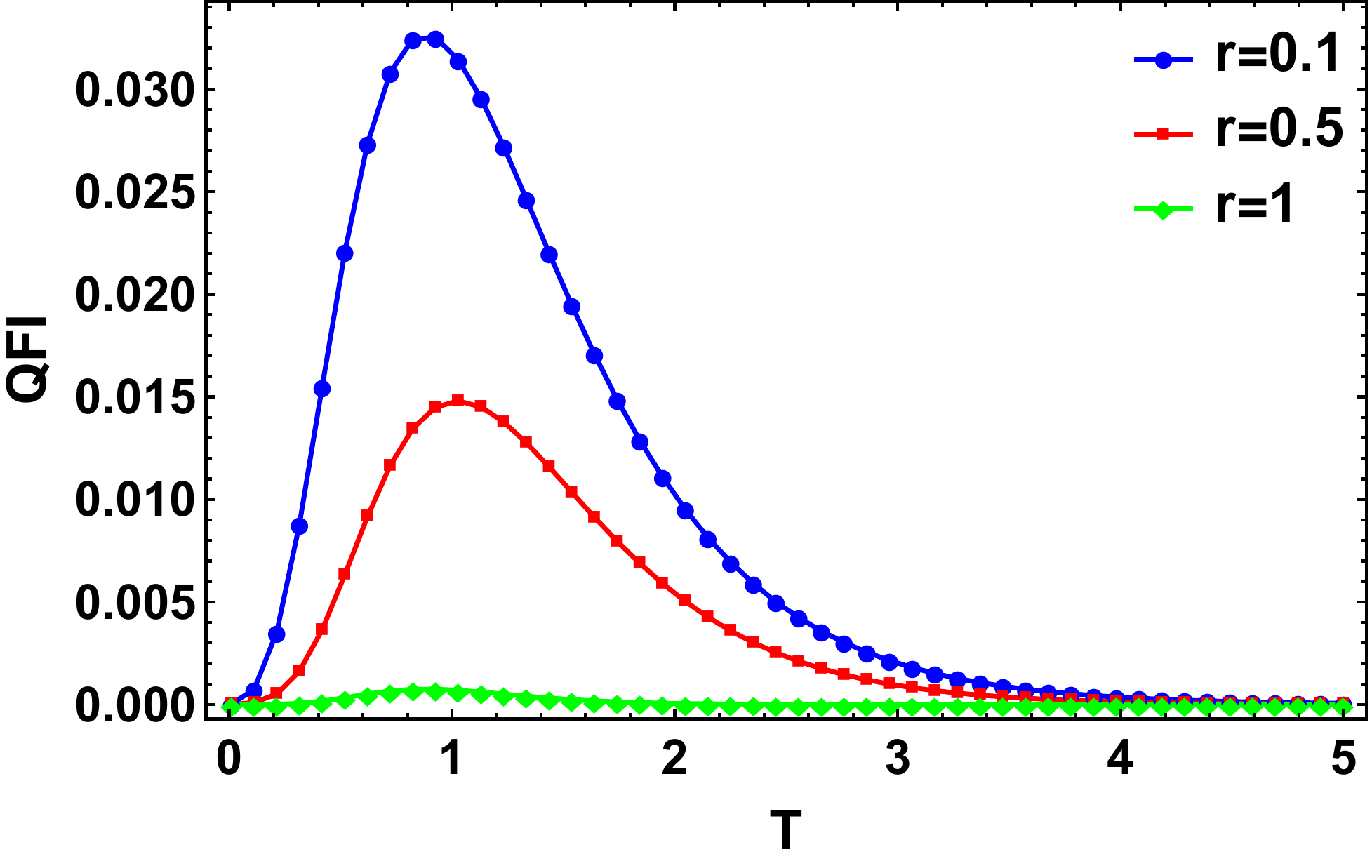}\label{Fig5a}}\hspace{0.4cm}
	\subfloat[]{
		\includegraphics[scale=0.25]{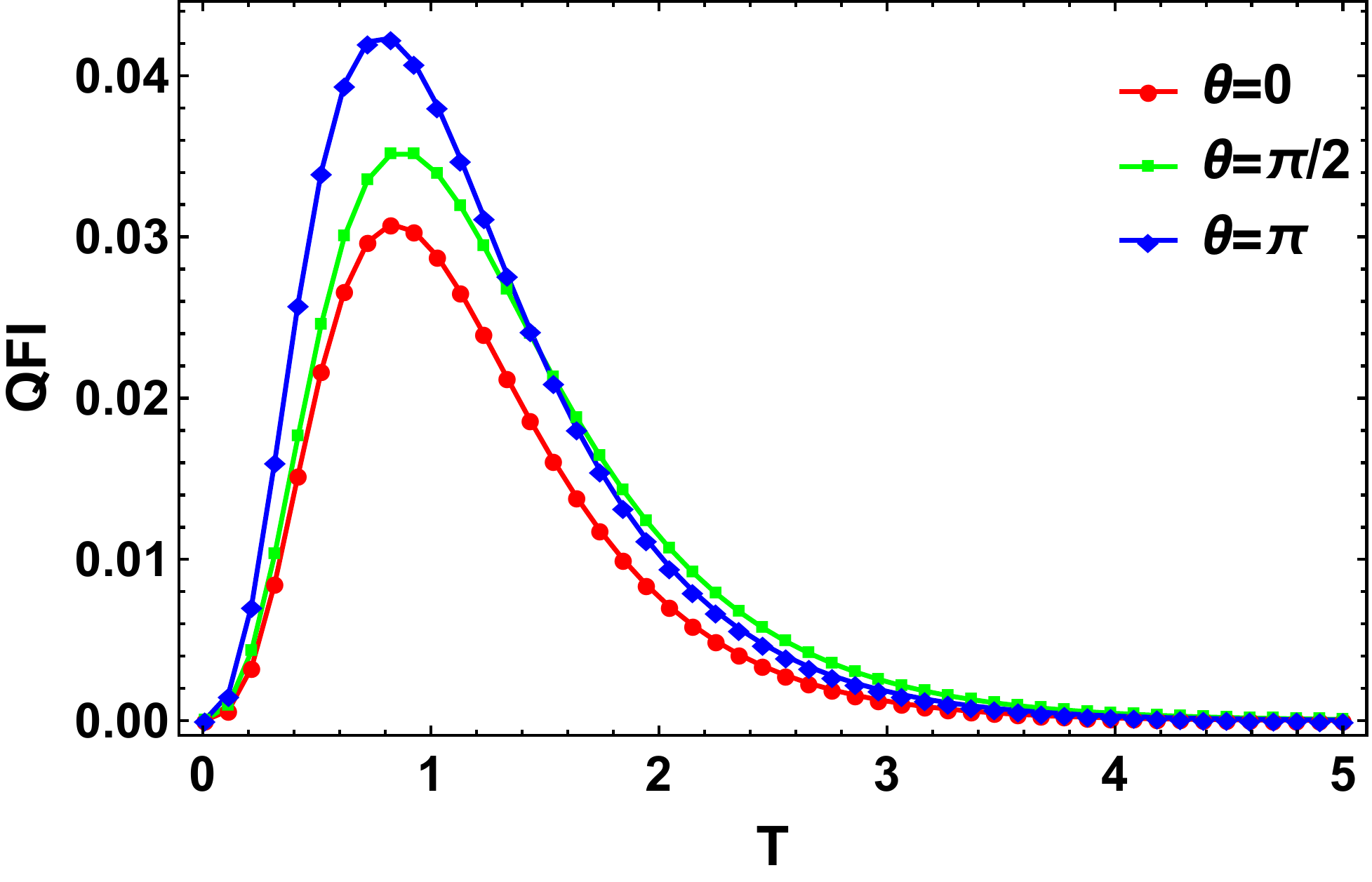}\label{Fig5b}}\\
	\subfloat[]{\includegraphics[scale=0.25]{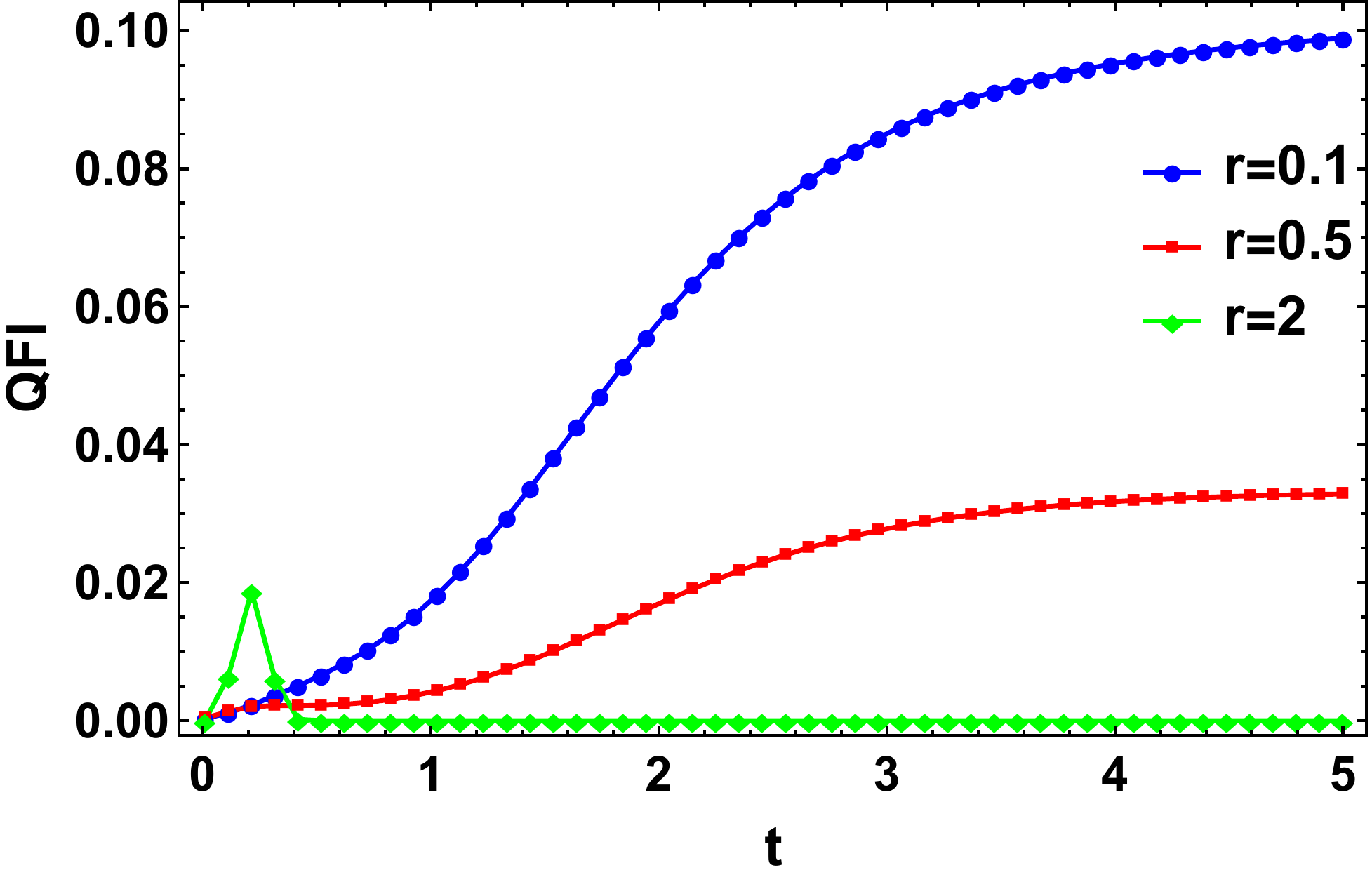}\label{Fig5c}}\hspace{0.4cm}
	\subfloat[]{
		\includegraphics[scale=0.25]{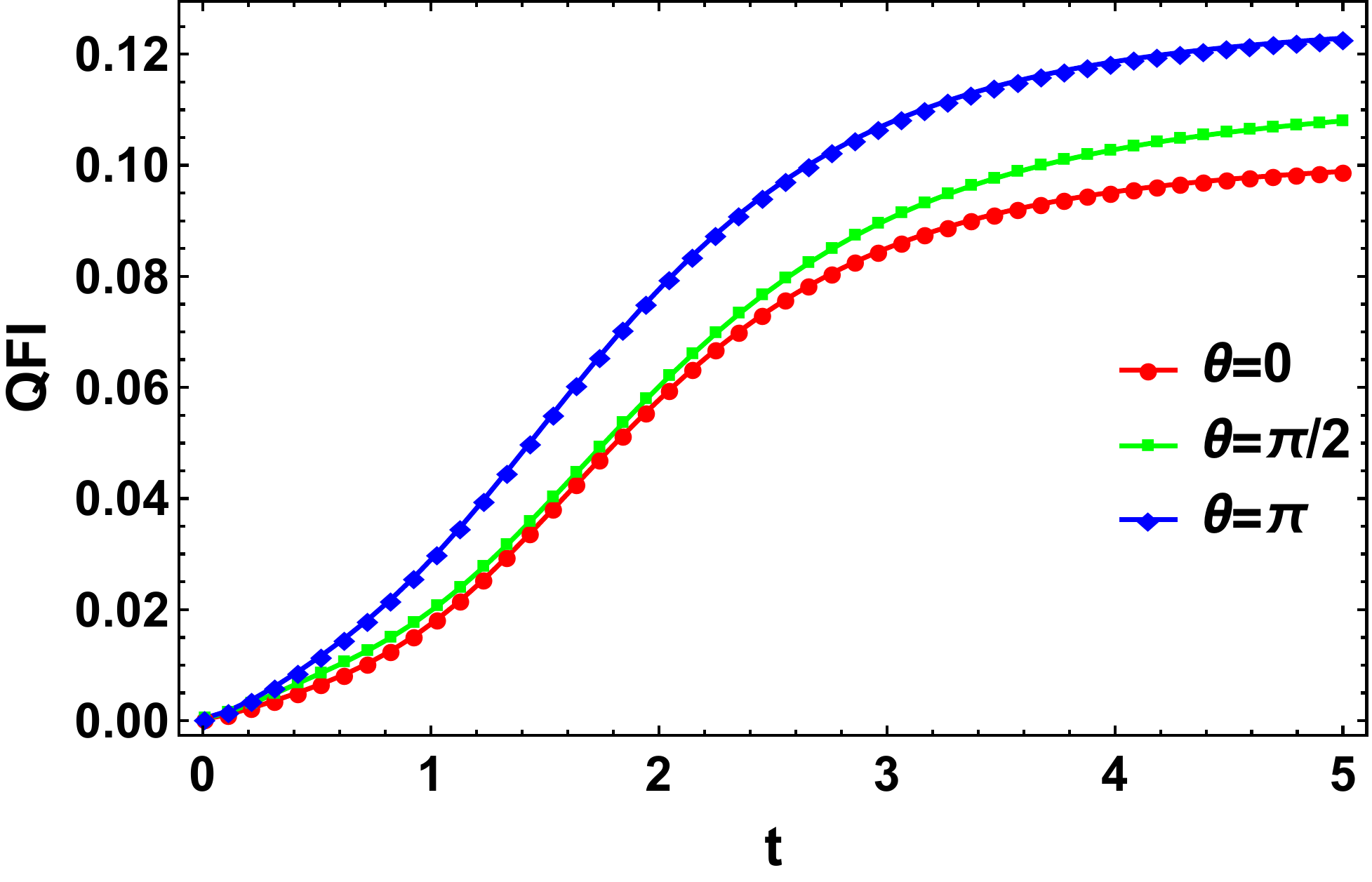}\label{Fig5d}}
	\caption{(Colour online) QFI verses temperature T (a,b) and QFI versus time t (c,d) for super-ohmic $(s=3)$ case.  Here we used (a) $t=1$, $\theta=1$ (b) $t=1$, $r=0.1$ (c) $T=0.5$, $\theta=1$ and (d) $T=0.5$, $r=0.1$.}\label{Fig5}
\end{figure}

In Figure \ref{Fig5}, QFI is plotted for the super-ohmic regime for $s=3$ with the rest parameters set to the same values as before. It can be identified from the figures that the peak values are further reduced and become negligibly small for the highest choice of $r$. Unlike its behavior in Figures \ref{fig1} and \ref{fig3} as a function of time, it saturates for each value of the parameters $r$ and $\theta$. The highly reduced maximum values of QFI against temperature reduces its utility in the super-ohmic case.

\begin{figure}[!]
	\centering
	\subfloat[]{
		\includegraphics[scale=0.3]{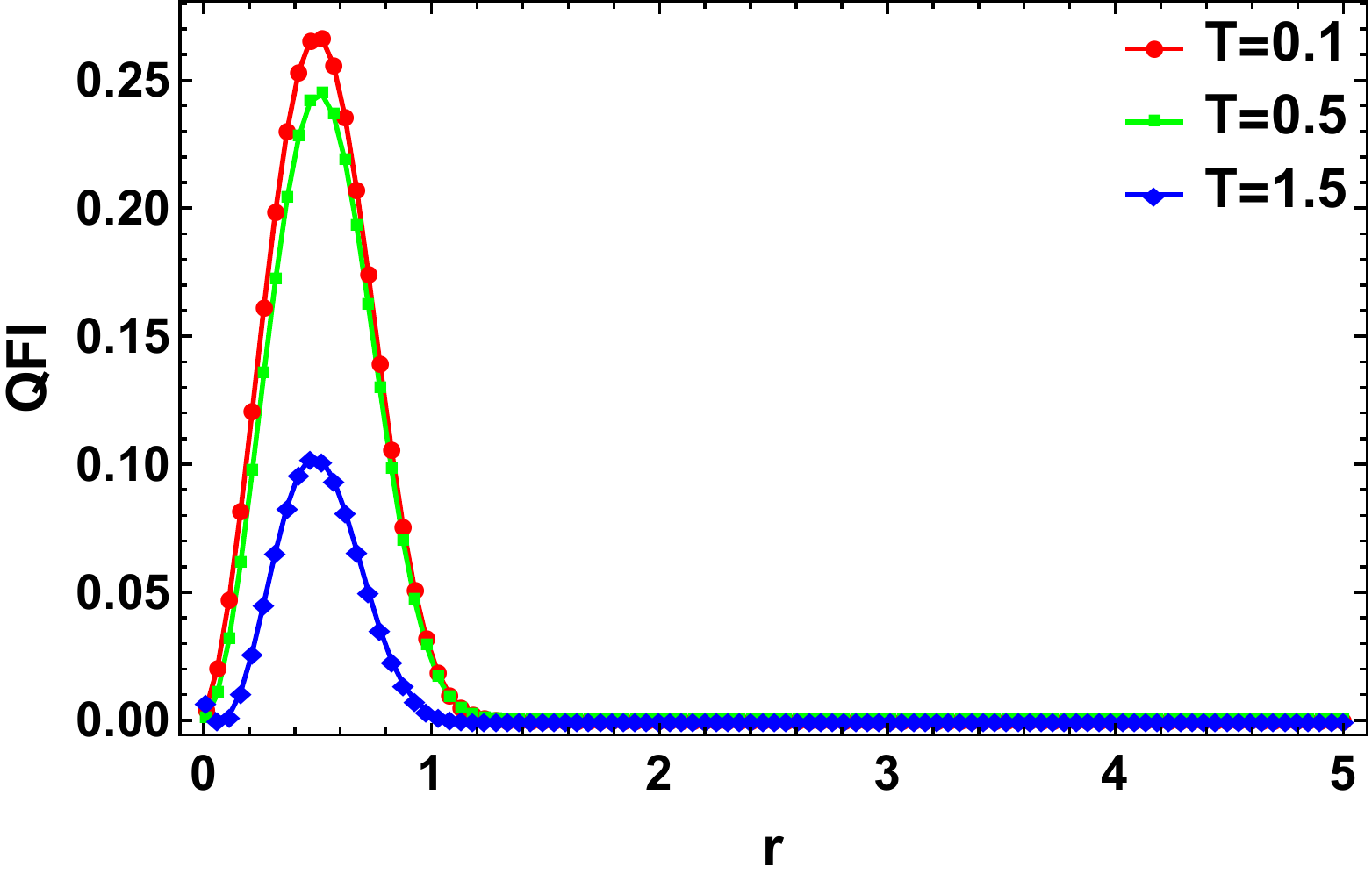}\label{Fig6a}}\hspace{0.4cm}
	\subfloat[]{
		\includegraphics[scale=0.3]{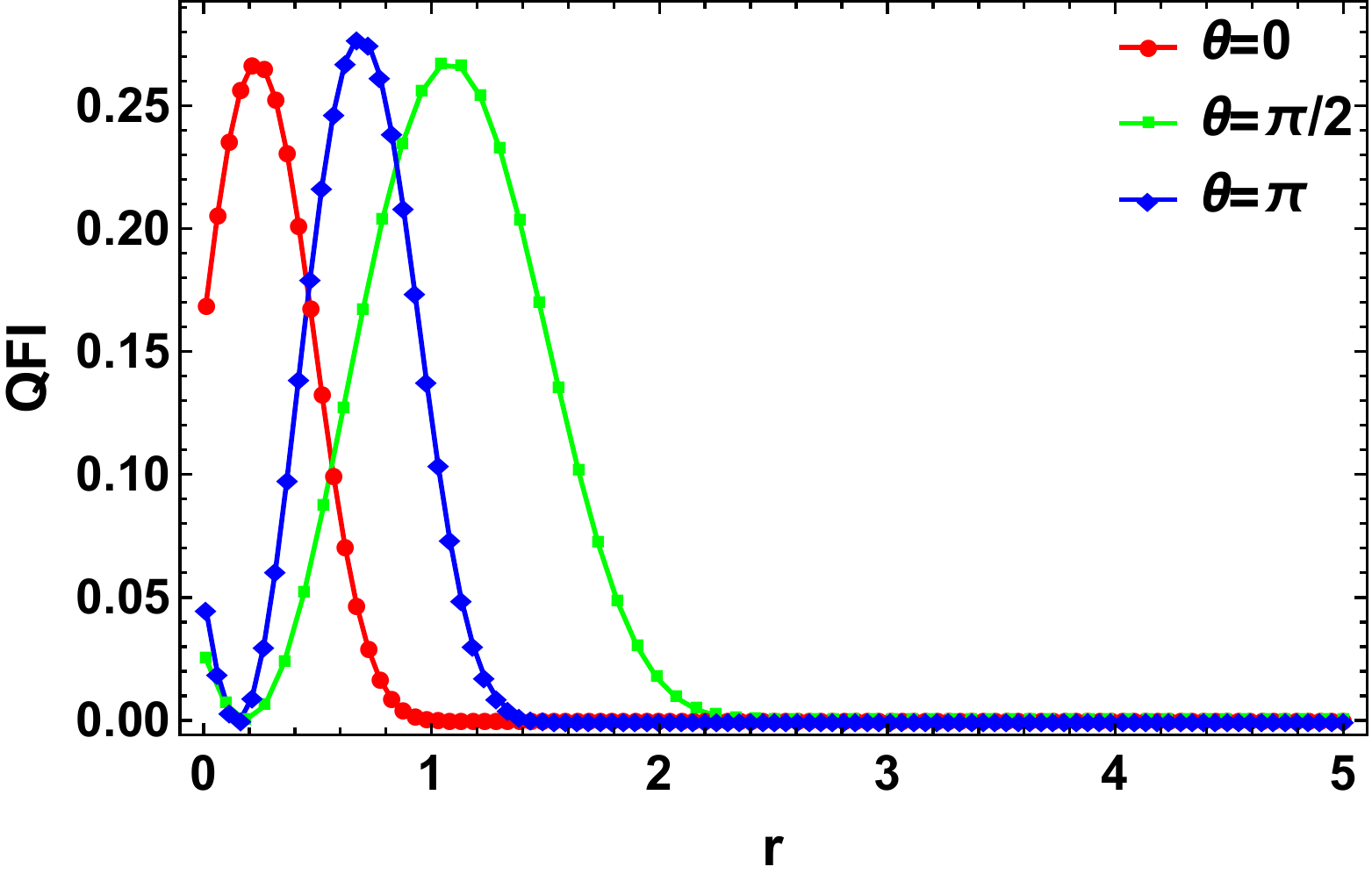}\label{Fig6b}}\\
	\subfloat[]{
		\includegraphics[scale=0.3]{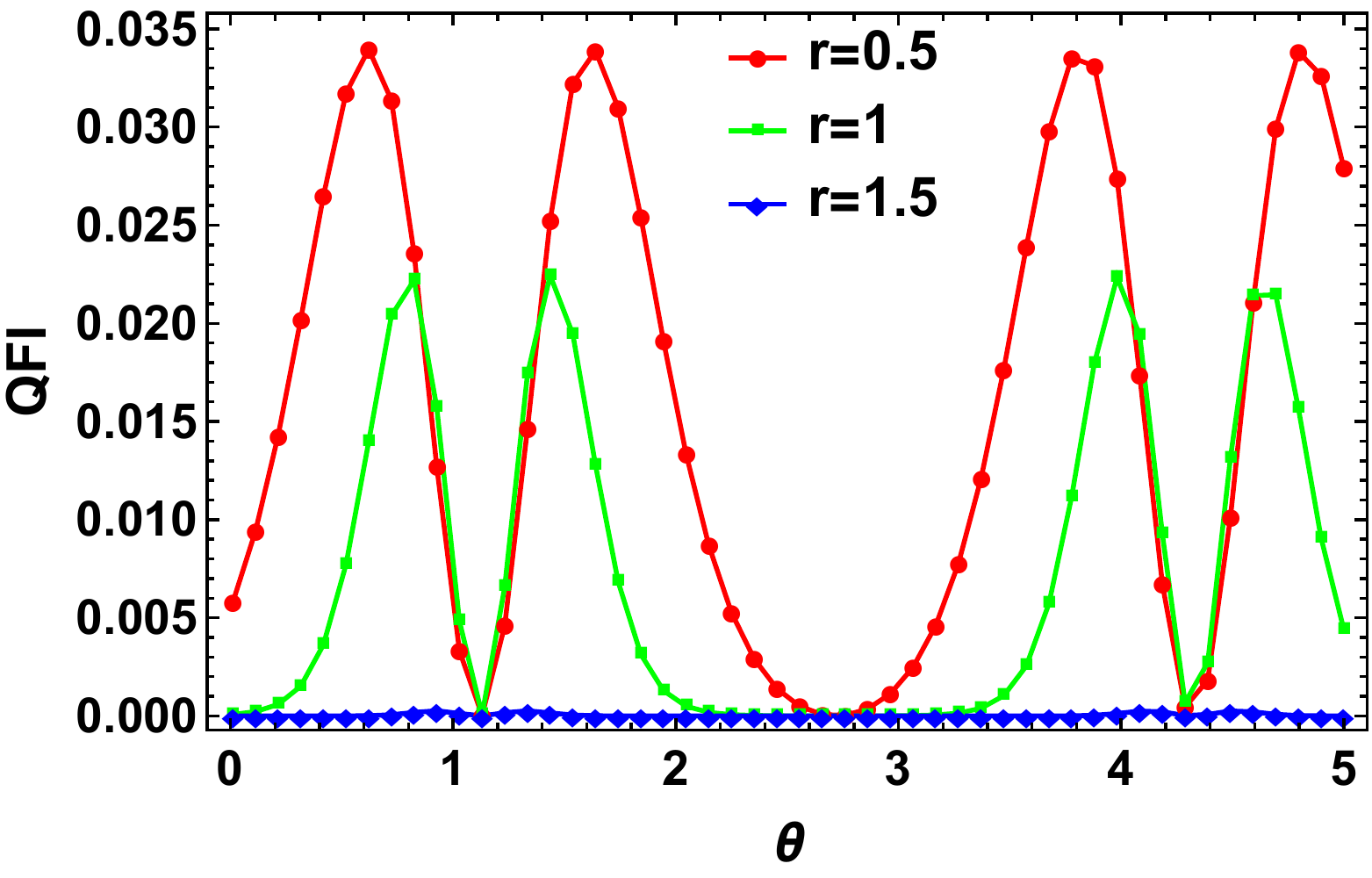}\label{Fig6c}}\hspace{0.4cm}
	\subfloat[]{
		\includegraphics[scale=0.3]{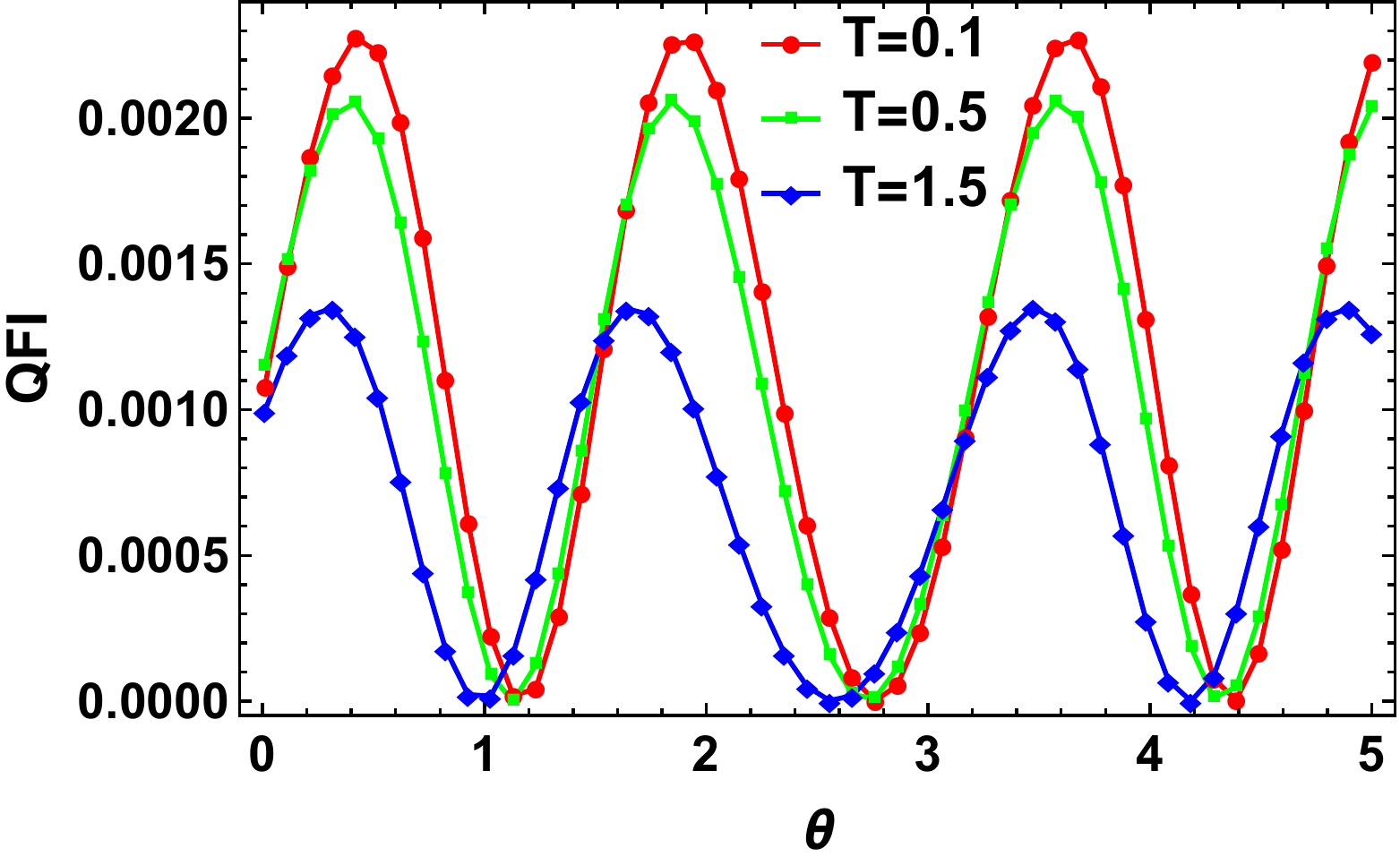}\label{Fig6d}}
	\caption{(Colour online) QFI verses  squeezing strength $r$ (a,b) and QFI versus phase parameter $\theta$ (c,d) for super-ohmic $(s=3)$ case. Here we used (a) $t=1$, $\theta=1$ (b) $t=1$, $T=0.5$ (c) $T=0.5$, $t=1$ and (d) $t=1$, $r=0.1$.}\label{Fig6}
\end{figure}

 QFI is plotted with respect to the parameters $r$ and $\theta$ for $s=3$ in Figure \ref{Fig6}. In super-ohmic regime, the peaks of QFI w.r.t $r$ are highest at low temperature but, the peak value of QFI is suppressed due to an increase in environmental temperature $T$ (Figure \ref{Fig6a}). The sharp peaks of QFI reduce the time needed to extract information about the parameter of the environment. However, the peak values of QFI almost remains the same for various values of $\theta$. In Figure (\ref{Fig6c}) and Figure (\ref{Fig6d}), we plotted QFI as a function of phase parameter $\theta$ for different choices of $r$ and $T$. As compared to ohmic and sub-ohmic regimes, i.e. Figures (\ref{Fig2c}) and (\ref{Fig4d}), the behavior of QFI as a function of $\theta$ is reversed shown in Figure (\ref{Fig6c}), as it is highly reduced with an increase in squeezing strength $r$. Similarly, a decrease in QFI with an increase in temperature can also be depicted in Figure (\ref{Fig6d}). This points out that the super-ohmic environment is not suitable for the estimation of phase parameter $\theta$.

\begin{figure}[!]
\centering
\subfloat[]{
	\includegraphics[scale=0.27]{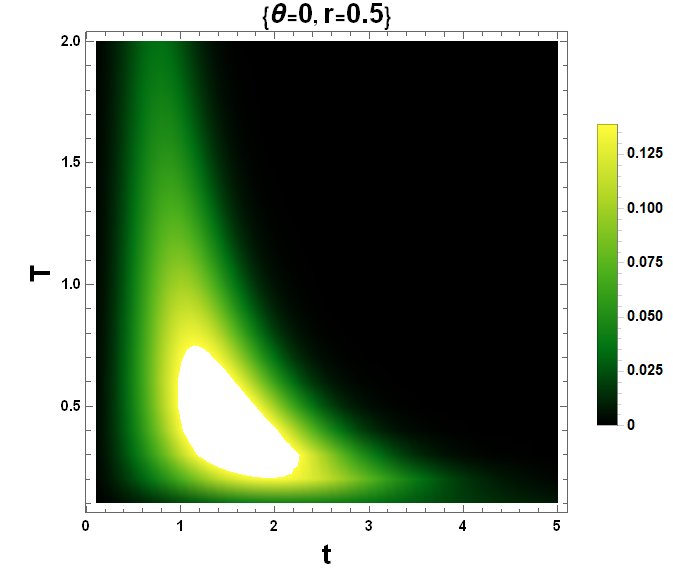}\label{Fig7a}}\hspace{0.4cm}
\subfloat[]{
	\includegraphics[scale=0.27]{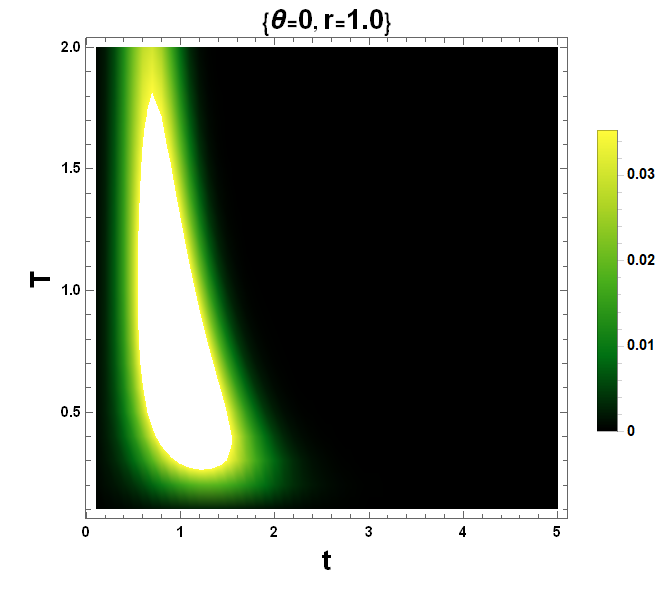}\label{Fig7b}}\\
\subfloat[]{
	\includegraphics[scale=0.27]{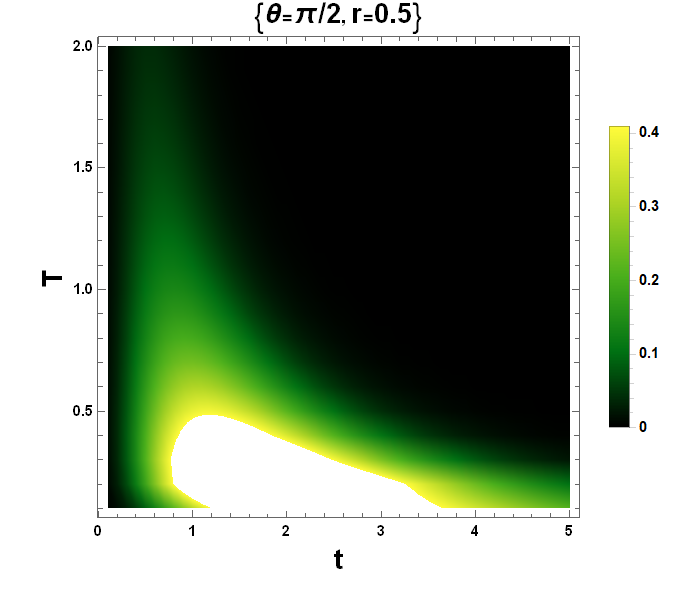}\label{Fig7c}}\hspace{0.4cm}
\subfloat[]{
	\includegraphics[scale=0.27]{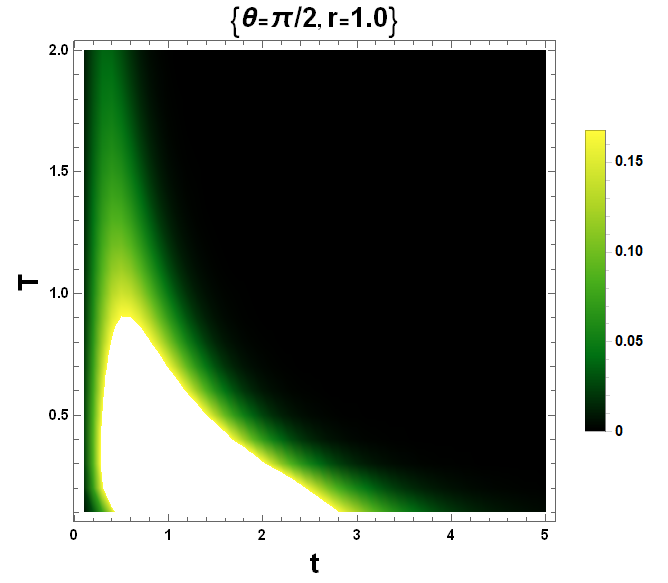}\label{Fig7d}}	
\caption{(Color online) Density plots exhibiting the variation of QFI with respect to the interaction time $t$ and temperature $T$ for selected values of $r$ and $\theta$ in sub-ohmic regime (s=0.5). The color bar at extreme right side shows the strength of QFI from its minimum to its maximum.}\label{Fig7}
\end{figure}

\begin{figure}[!]
	\centering
	\subfloat[]{
		\includegraphics[scale=0.27]{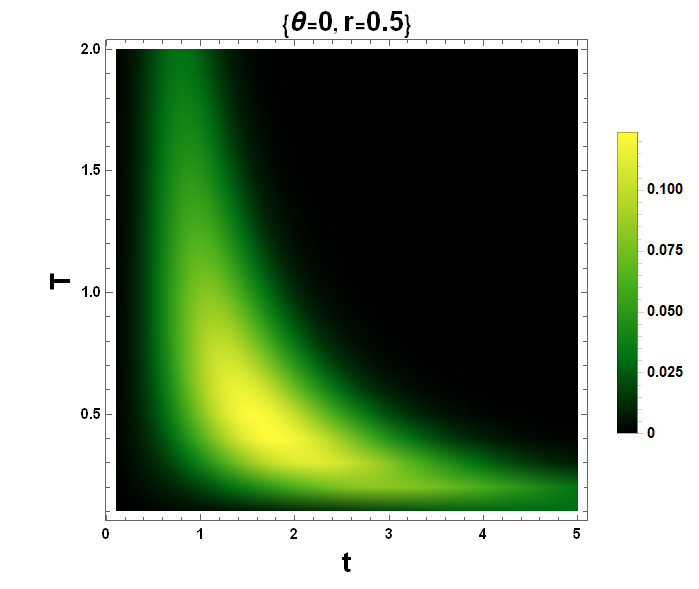}\label{Fig8a}}\hspace{0.4cm}
	\subfloat[]{
		\includegraphics[scale=0.27]{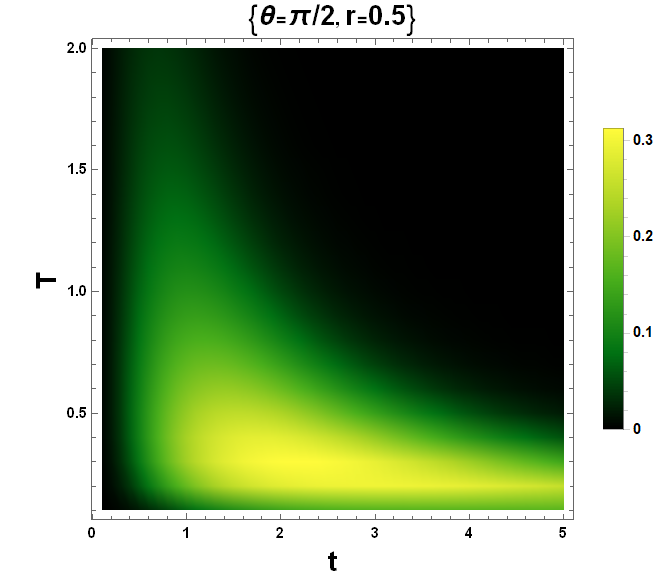}\label{Fig8b}}\\
	\subfloat[]{
		\includegraphics[scale=0.27]{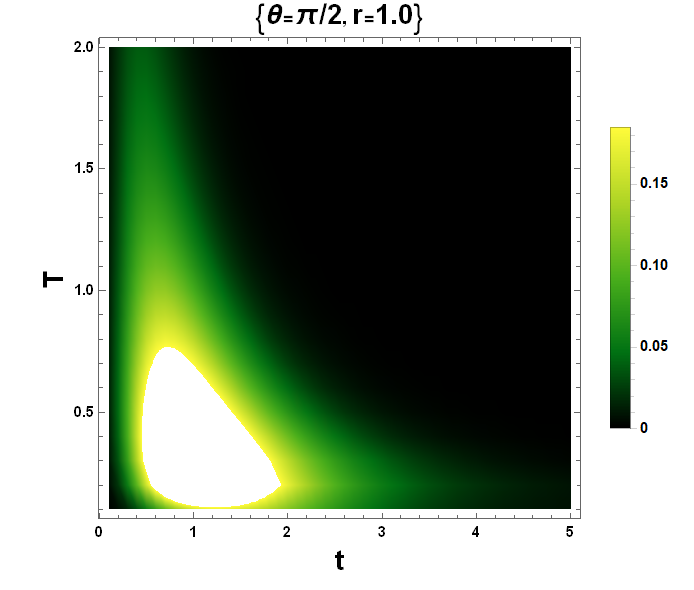}\label{Fig8c}}\hspace{0.4cm}
	\subfloat[]{
		\includegraphics[scale=0.27]{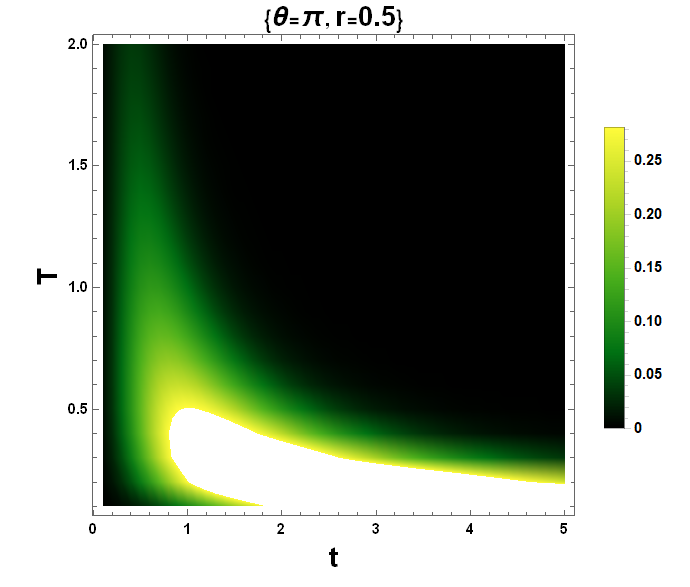}\label{Fig8d}}	
	\caption{(Color online) Density plots exhibiting the variation of QFI with respect to the interaction time $t$ and temperature $T$ for selected values of parameter $r$ and $\theta$ in ohmic regime (s=1). The color bar at extreme right side shows the strength of QFI from its minimum to its maximum.}\label{Fig8}
	
\end{figure}

\begin{figure}[!]
	\centering
	\subfloat[]{
		\includegraphics[scale=0.27]{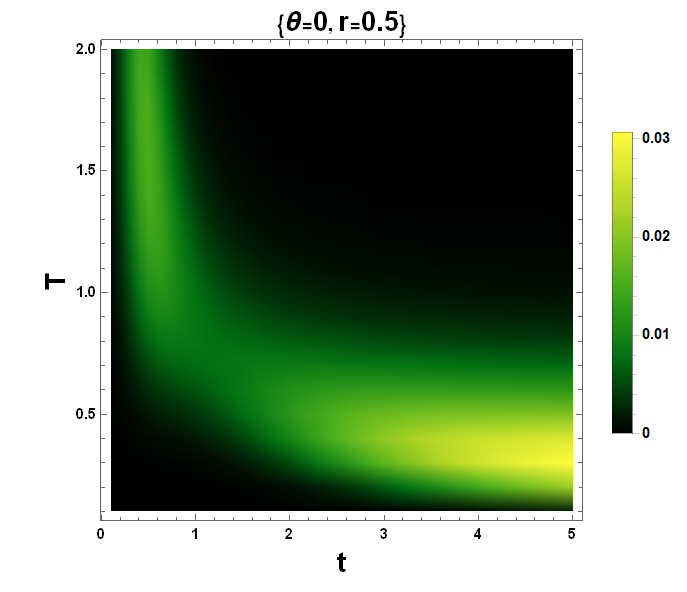}\label{Fig9a}}\hspace{0.4cm}
	\subfloat[]{
		\includegraphics[scale=0.27]{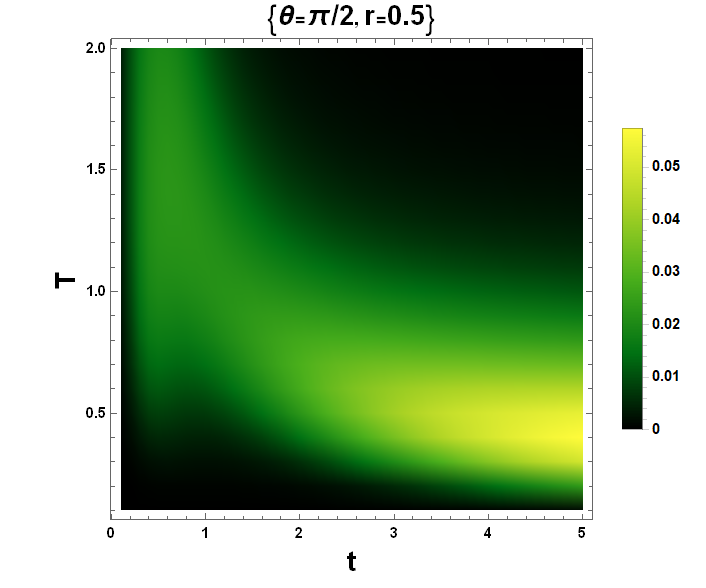}\label{Fig9b}}\\
	\subfloat[]{
		\includegraphics[scale=0.27]{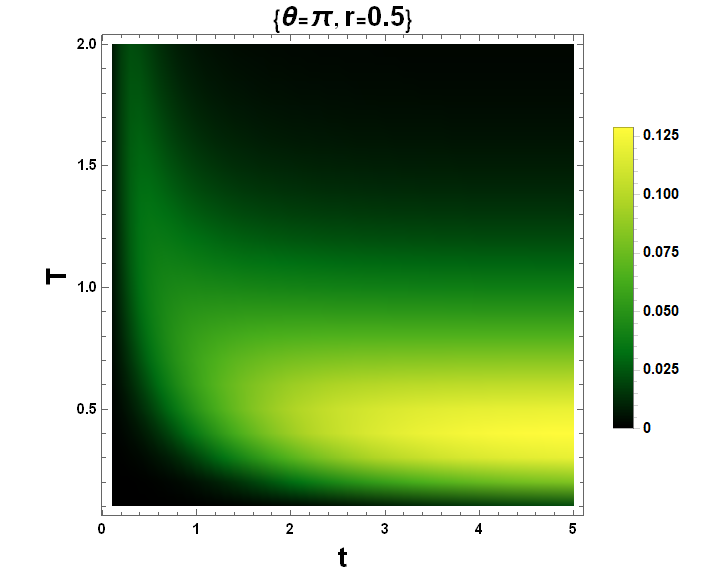}\label{Fig9c}}\hspace{0.4cm}
	\subfloat[]{
		\includegraphics[scale=0.27]{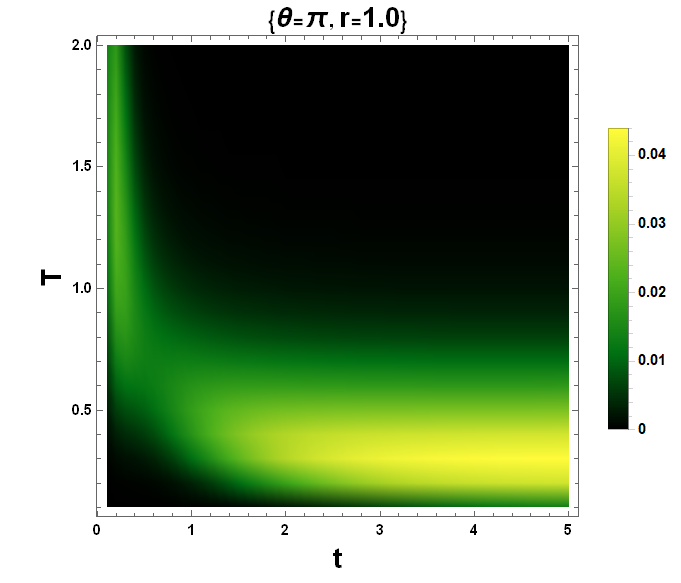}\label{Fig9d}}	
	\caption{(Color online) Density plots exhibiting the variation of QFI with respect to the interaction time $t$ and temperature $T$ for selected values of parameter $r$ and $\theta$ in super-ohmic regime (s=3). The color bar at extreme right side shows the strength of QFI from its minimum to its maximum.}\label{Fig9}
\end{figure}

 To get a deeper and clear insight into the behavior of QFI, we plotted QFI as a function of time and temperature at different value of parameter $r$ and $\theta$ for three dynamics of the environment, i.e sub-ohmic, ohmic and super-ohmic given in Figures \ref{Fig7}, \ref{Fig8} and \ref{Fig9}. The dependence of QFI on $r$ and $\theta$ with the variation of time and temperature can be observed from the graphs. In Figures \ref{Fig7} and \ref{Fig8}, one can easily interpret that for certain values of $t$ and $T$, the value of QFI is maximum. An increase in temperature reduces the maximum value of QFI. The whiter regions in these plots correspond to the maximum values of QFI. In super-ohmic case, as shown in Figure \ref{Fig9}, one cannot notice the exact value of maximum QFI as a function of time and temperature for different values of $r$ and $\theta$. The maximum value of QFI is only found at high temperature, while at low temperature, for some specific time interval, a maximum value of QFI cannot be observed.
\begin{figure}[!]
	\centering
	\subfloat[]{
		\includegraphics[scale=0.3]{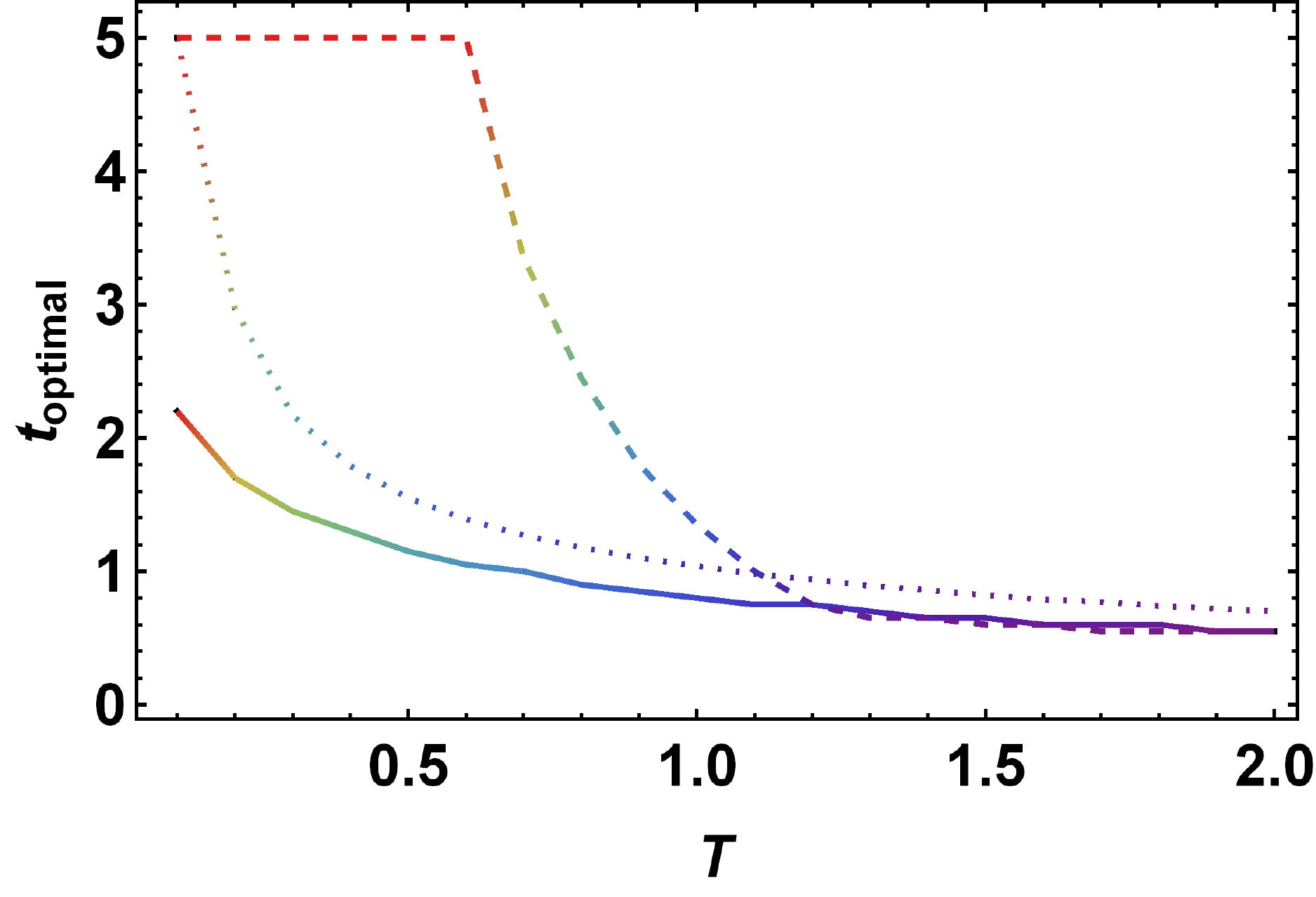}}\hspace{0.4cm}
	\caption{(Color online) The plot of optimal time when maximal QFI is attained as a function of temperature $T$ where solid line is for sub-ohmic regime ($s=0.5$), dotted line is for ohmic regime ($s=1$) and dashed line is for super-ohmic regime ($s=3$) and, we used $\theta=\pi/2$ and $r=0.5$.}\label{Fig10}
\end{figure}
Figure \ref{Fig10} denotes the results of optimization for optimal time, i.e. $t_{optimal}$, the time at which maximal QFI is achieved as a function of temperature for three different values of ohmicity parameter $r$ with $\theta=\pi/2$, $r=0.5$. The plot for the sub-ohmic regime (dashed line) depicts that at low temperature, the time needed to obtain information from the environment is larger and the time reduces gradually as the temperature increases. This reveals that the decoherence is not strong at low temperature, which in turns increases $t_{optimal}$. In the ohmic regime (dotted line), the optimal time is larger as compared to the sub-ohmic case. In comparison to the super-ohmic regime (dashed line), for some specific low-temperature range, the optimal time is constant and, it reduces quickly to a lower value and after $T=1.2$, it reaches a constant value. These results manifest a robust dependence of QFI on the ohmicity parameter in lieu of $r$ and $\theta$.
\section{Conclusion}\label{secIV}
In this paper, we have studied the effect of various parameters of squeezed thermal environment on the dynamics of QFI by using a qubit as a probe. We have analytically found the optimal initial qubit state that maximizes QFI. We have further investigated the optimal temperature and optimal time of interaction through numerical simulation of the analytical result for QFI. The dynamics of QFI in the three different regimes of environment, namely, sub-ohmic, ohmic, and super-ohmic, are compared with each other. In each case, the qualitative behavior of QFI as a function of temperature is similar, it increases initially reaching a maximum followed by a gradual monotonic decay and reach a vanishing value in the asymptotic limit. However, quantitatively the dynamics strongly depend on the ohmicity parameter, characterizing the three regimes of the environment. The existence of optimal temperature and optimal initial time in each regime means that their values can precisely be estimated through the estimation procedures, which may be beneficial in quantum metrology. Furthermore, we have also studied the effect of different parameters on the precise estimation of squeezing amplitude and phase of the squeezed thermal environment. It is concluded from our results that, Sub-ohmic and ohmic regimes can provide the possibility for precise estimation of the temperature and squeezing phase however, the low-temperature regime befits the estimation of squeezing amplitude for all the three different dynamics of the environment.

\end{document}